\documentclass[12pt,preprint]{aastex}

\newcommand{\fmi}{\, \rm{fm}^{-1}}
\newcommand{\fmiq}{\, \rm{fm}^{-3}}

\newcommand{\mev}{\, \rm{MeV}}

\begin{document}

\title{Equation of state and neutron star properties 
constrained by nuclear physics and observation}

\author{K.\ Hebeler,$^1$ J.\ M.\ Lattimer,$^{2}$ C.\ J.\ Pethick,$^{3,4}$
and A.\ Schwenk$\,^{5,6}$}
\affil{$^1$Department of Physics,
The Ohio State University, Columbus, OH 43210, USA \\
$^{2}$Department of Physics and Astronomy, \\
Stony Brook University, Stony Brook, NY 11794-3800, USA \\
$^3$The Niels Bohr International Academy,
The Niels Bohr Institute, \\ University of Copenhagen, Blegdamsvej 17,
DK-2100 Copenhagen \O, Denmark \\
$^4$NORDITA, KTH Royal Institute of Technology and Stockholm University,
Roslagstullsbacken 23, SE-10691 Stockholm, Sweden \\
$^5$ExtreMe Matter Institute EMMI,
GSI Helmholtzzentrum f\"ur Schwerionenforschung~GmbH,
D-64291 Darmstadt, Germany \\
$^6$Institut f\"ur Kernphysik,
Technische Universit\"at Darmstadt, D-64289 Darmstadt, Germany}

\begin{abstract}
Microscopic calculations of neutron matter based on nuclear
interactions derived from chiral effective field theory, combined with
the recent observation of a $1.97 \pm 0.04 \, M_{\odot}$ neutron star,
constrain the equation of state of neutron-rich matter at sub-
and supranuclear densities. We discuss in detail the allowed equations
of state and the impact of our results on the structure of neutron
stars, the crust-core transition density, and the nuclear symmetry
energy. In particular, we show that the predicted range for neutron
star radii is robust. For use in astrophysical simulations, we provide
detailed numerical tables for a representative set of equations of
state consistent with these constraints.
\end{abstract}


\maketitle

\section{Introduction}
\label{sec:intro}

Neutron stars, apart from being systems for investigating such diverse
topics as theories of gravity and the interstellar medium, are unique
laboratories for studying matter at high densities. In neutron stars,
matter ranges from nuclei embedded in a sea of electrons at low
densities in the outer crust, to increasingly neutron-rich structures
in the inner crust, to the extremely neutron-rich uniform matter in
the outer core, and possibly exotic states of matter at high densities
in the inner core~\citep{nstar_book,Lattimer}.  The theoretical
understanding of nuclear matter and atomic nuclei over such a range of
densities and isospin asymmetry is a current frontier in nuclear
science. The equation of state (EOS) of dense matter is also a key
ingredient in modeling neutron star and black hole formation and, in
particular, the gravitational wave signal from mergers of binary
neutron stars~\citep{Anderson,Bauswein,Bauswein2} and neutron
star-black hole mergers~\citep{Lackey} is sensitive to
it. Consequently, future searches with advanced LIGO and LISA are
expected to provide information about the high-density EOS.

In the past, the EOSs of dense matter most commonly used in
astrophysical simulations have been based on phenomenological nuclear
interactions [for a review see, for example, \citet{HP}]. Two-nucleon
(NN) interactions are usually constructed to fit NN scattering data at
low energies. In addition, three-nucleon (3N) interactions were
introduced because, first, they exist on theoretical grounds and,
second, NN interactions alone cannot reproduce the properties of
nuclei and nuclear matter.  The many-body problem based on those
interactions is very challenging due to strongly repulsive
forces at small relative distances, which lead to highly correlated
wave functions and require nonperturbative many-body methods.

In recent years, the development of chiral effective field theory
(EFT), following the pioneering work of~\citet{Weinberg1,Weinberg2},
has provided the framework for a systematic expansion for nuclear
forces at low momenta, where nucleons interact by pion exchanges and
short-range contact interactions whose parameters can be fixed on the
basis of two- and few-body observables~\citep{RMP}. Chiral EFT
explains the hierarchy of two-, three-, and weaker higher-body forces
and provides estimates of the theoretical uncertainties. In a recent
Letter~\citep{Kai}, we have shown that microscopic calculations based
on chiral EFT interactions constrain the properties of neutron-rich
matter up to nuclear saturation density to a high degree. On the
basis of laboratory experiments and theory, our knowledge of the EOS
at densities greater than $1-2$ times the saturation density is
limited. However, information may be obtained from measurements of
neutron star masses. In particular, the recent discovery of a neutron
star with a precisely determined mass of $1.97 \pm 0.04 \,
M_{\odot}$~\citep{Demorest}, the heaviest to date, is extremely
important, as it rules out a large number of EOSs based on exotic
degrees of freedom like hyperons or deconfined quarks. Such
constituents soften the EOS at high densities and, without
fine-tuning, are generally incompatible with a neutron star of such a
large mass. Further information may be obtained, for example, from
modeling X-ray bursts and quiescent low-mass X-ray
binaries~\citep{Ozel,Steiner1,Ozel12,Steiner2,Guever13}, but the neutron star
properties deduced are more model-dependent than the direct mass
constraint from a very heavy neutron star.

By extending our microscopic results for the EOS at low densities in a
general way to higher densities, we showed in~\citet{Kai} that it is
possible to derive systematic constraints on the EOS and on the radii
of neutron stars. The high-density extensions we used were only
constrained by causality and by the heaviest observed neutron star at
that time which has a mass of $1.65 \, M_{\odot}$. In this paper, we
present details of these calculations and generalize and improve our
approach in several ways: First, we require that the EOS is consistent
with the observation of a $1.97 \, M_{\odot}$ neutron
star~\citep{Demorest}. To investigate the sensitivity to the possible 
future discovery of neutron stars of higher mass, we also 
consider a second case where the EOS supports a
neutron star of mass $2.4 \, M_{\odot}$.  Second, we generalize
the microscopic neutron matter calculations. Our previous results were
based on renormalization-group-evolved NN interactions plus the
leading 3N interactions from chiral EFT. The renormalization-group
evolution improves the many-body convergence~\citep{PPNP}, but
introduces uncertainties because the evolution was limited to NN
forces. Here, we show that calculations based on unevolved chiral EFT
interactions are in good agreement with the previous calculations (see
Section~\ref{sec:neutron_matter}). Third, we improve the way beta
equilibrium is incorporated (see Section~\ref{sec:beta}) and include
explicitly the crust EOS below the crust-core transition density,
which we calculate in Section~\ref{sec:crust-core}. Finally, we
generalize the piecewise polytropic extensions of the EOS to higher
densities by allowing more density regions and also refine the step
size of the variation of the polytropic parameters (see
Section~\ref{sec:gen_ext}). We present our results for the nuclear EOS
and the structure of neutron stars in Section~\ref{sec:constraints}.
This shows that the constraints are robust and not significantly
altered by the generalizations and improvements of the microscopic
calculations and EOS extensions. For use in astrophysical simulations,
we construct in Section~\ref{sec:repEOS} three representative EOSs
consistent with the constraints from nuclear physics and observations
and provide numerical data in Appendix~\ref{tables}.

\section{Neutron matter}
\label{sec:neutron_matter}

Our microscopic neutron matter calculations are based on chiral NN and
3N interactions. Neutron matter presents a unique system in chiral EFT
because only the long-range two-pion-exchange parts of the leading 3N
interactions contribute~\citep{nm}. This is because three neutrons
cannot interact via point-like S-wave interactions due to the Pauli
principle. Moreover, the leading one-pion-exchange 3N interaction does
not contribute in neutron systems because of the particular
spin-momentum structure of this interaction.

The leading chiral 3N forces have been shown to give important
contributions to the nuclear EOS and to properties of
nuclei~\citep{RMP3N}. In particular, saturation of symmetric nuclear
matter is driven by 3N forces~\citep{nucmatt}. While 3N contributions
to the neutron matter EOS are smaller in size than for nuclear matter,
they are still significant and crucial for predictions of observables
like the nuclear symmetry energy [see \citet{nm} and Section~\ref{sec:beta}].
In this work, we include only the leading 3N forces, because it is
presently possible to include only these beyond the Hartree-Fock
level~\citep{N3LO}.

For calculations of the neutron matter EOS, the theoretical
uncertainty is dominated by the uncertainties in the low-energy
couplings $c_1$ and $c_3$, which determine the two-pion-exchange
three-body interactions between neutrons, not by the many-body
approximations~\citep{nm}. These low-energy couplings relate $\pi$N,
NN and 3N interactions. Their determination from $\pi$N scattering is
within uncertainties consistent with the extraction from NN
scattering, leading to $c_1 = - (0.7 - 1.4) \, {\rm GeV}^{-1}$ and
$c_3 = -(3.2 - 5.7) \, {\rm GeV}^{-1}$~\citep{EM,EGM,Rentmeester,%
Meissner_private}. Beyond the leading 3N forces, part of the
subleading 3N forces are simple shifts of $c_1$ and
$c_3$~\citep{RMP}. Therefore, as in \citet{Kai}, we take $c_1 = - (0.7
- 1.4) \, {\rm GeV}^{-1}$ and $c_3 = -(2.2 - 4.8) \, {\rm GeV}^{-1}$
for our calculations.

In general, nuclear forces depend on an intrinsic resolution scale
$\Lambda$. Consequently, the nuclear Hamiltonian can be written in
the form
\begin{equation}
H(\Lambda) = T + V_{\rm NN}(\Lambda) + V_{\rm 3N}(\Lambda) 
+ V_{\rm 4N}(\Lambda) + \ldots \,,
\label{eq:H}
\end{equation}
where $T$ denotes the kinetic energy, $V_{\rm NN}$ the NN
interactions, $V_{\rm 3N}$ the 3N interactions, etc. The
renormalization group provides a powerful tool to systematically
change the scale $\Lambda$, while low-energy observables are
preserved~\citep{PPNP}. The evolution to low momentum scales improves
the many-body convergence due to a decoupling of low and high momenta
in the Hamiltonian~\citep{PPNP}. In general, the renormalization-group
evolution changes all terms in Equation~(\ref{eq:H}). The consistent
evolution of 3N interactions in momentum space is a complex task and
has been achieved only recently~\citep{3N_evolution_mom}.

\begin{figure}[t]
\begin{center}
\includegraphics[scale=0.9,clip=]{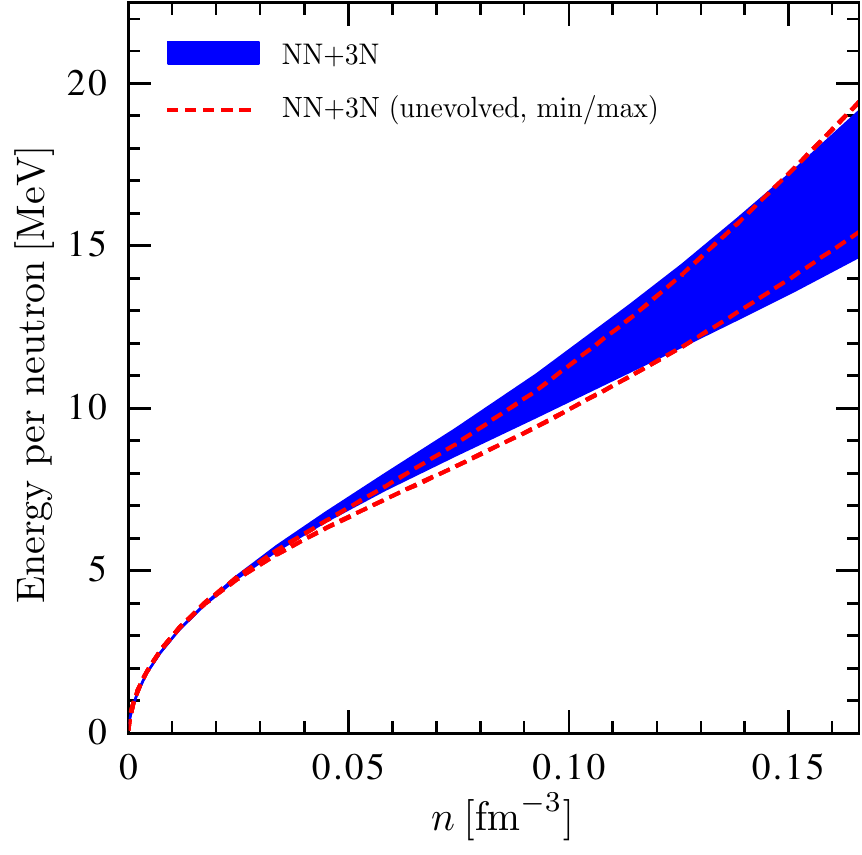}
\end{center}
\caption{(Color online) Energy per particle of neutron matter as a
function of density $n$. The blue band is based on chiral NN and 3N
interactions with a renormalization-group evolution to improve the
many-body convergence. The range of the band is mainly due to 
uncertainties in 3N forces~\citep{nm}. The dashed red lines present
the range without the renormalization-group evolution.\label{EN_nobeta}}
\end{figure}

\begin{figure}[t]
\begin{center}
\includegraphics[scale=0.9,clip]{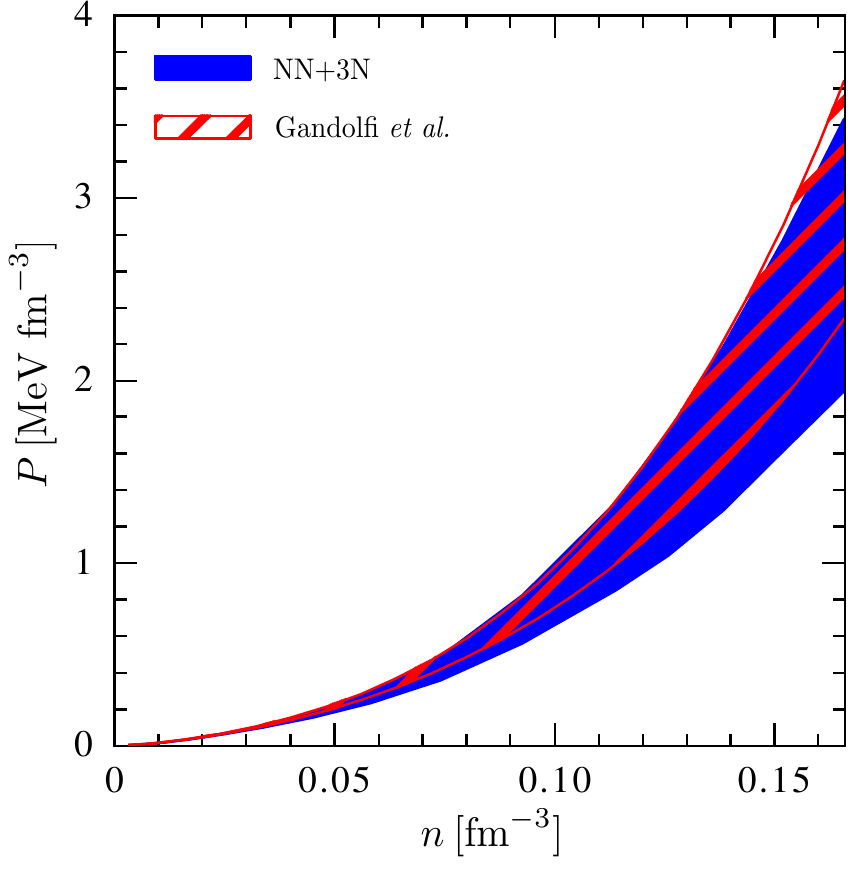}
\end{center}
\caption{(Color online) Pressure $P$ of neutron matter as a function of
density $n$. The blue band is as in Figure~\ref{EN_nobeta} based on
chiral NN and 3N interactions. For comparison, the shaded red band
shows the Quantum Monte Carlo results of \citet{Gandolfi} with
3N forces fitted to a symmetry energy of $32.0-35.1 
\mev$ [see Table~I of \citet{Gandolfi}].\label{p_nobeta}}
\end{figure}

We start from the chiral N$^3$LO NN potential with $\Lambda = 500
\mev$ of \citet{EM} and use the renormalization group to evolve this
NN potential to low-momentum scales $\Lambda = 1.8 - 2.8
\fmi$~\citep{smooth,NN_evolution}. At these scales, NN interactions
derived from different initial potentials are very
similar~\citep{Vlowk}. This universality can be attributed to common
long-range pion physics and phase-shift equivalence. As a consequence,
results of many-body calculations are rather insensitive to the
particular choice of the initial NN interaction. Because the leading
chiral 3N forces are of long-range character in neutron matter, they
are expected to be to a good approximation invariant 
under the renormalization-group evolution
for these cutoffs. Therefore, we use the leading chiral 3N forces
determined by $c_1$ and $c_3$ also with low-momentum cutoffs.

For low-momentum cutoffs, the EOS of neutron matter and nuclear matter
can be calculated with theoretical uncertainties in a perturbative
expansion in which the Hartree-Fock approximation is the first-order
term~\citep{nm,nucmatt}. Our neutron matter results also include
second-order corrections~[for calculational details, see~\citet{nm}].
For low-momentum cutoffs $1.8 \fmi \leqslant \Lambda \leqslant 2.8
\fmi$, we have checked that contributions from third-order
particle-particle diagrams give only small contributions to the energy
per particle, about $25 \, {\rm keV}$ for $\Lambda=1.8 \fmi$ and $300
\, {\rm keV}$ for $\Lambda=2.8 \fmi$ at the saturation density $n_0 =
0.16 \fmiq$. In addition, we find that our second-order results are
independent of the resolution scale to a very good approximation, and
at the saturation density the maximal variation is about 1~MeV. We
also find that the results are insensitive to the single-particle
spectrum used. All these findings indicate that neutron matter is
perturbative for low-momentum interactions, and show that the
theoretical uncertainties of the many-body calculation are
small~\citep{nm}.

Figure~\ref{EN_nobeta} shows our results for the energy per particle
of neutron matter up to the saturation density (using a Hartree-Fock
spectrum). The blue band is based on chiral NN and 3N interactions,
with a renormalization-group evolution for NN interactions.  The width
of the band is due mainly to the uncertainties of $c_1$ and $c_3$ in
3N forces~\citep{nm}. For comparison, the dashed red lines present the
range based on unevolved chiral NN interactions plus the same leading
chiral 3N interactions. The remarkable agreement indicates that
neutron matter is, to a good approximation, also perturbative for
chiral NN interactions [for a detailed study, see \citet{N3LO_long}].
We also explicitly checked the size of the contributions at
third-order in the many-body expansion. As expected, they are larger
for unevolved chiral NN interactions due to the stronger coupling
between low and high momenta in the Hamiltonian. However, they are
still significantly smaller than the second-order
contributions~\citep{N3LO_long}.

In Figure~\ref{p_nobeta} we present the uncertainty band for the
pressure of neutron matter based on chiral NN and 3N interactions
(with a renormalization-group evolution for NN interactions).  For
comparison, we also show the Quantum Monte Carlo results of
\citet{Gandolfi} (shaded red band) based on the phenomenological
Argonne $v_{18}$ NN potential plus 3N forces fitted to a symmetry
energy of $32.0-35.1 \mev$ [see Table~I of \citet{Gandolfi}]. The
agreement of the results is remarkable, given that the Hamiltonian and
the many-body methods are completely different.

\section{Asymmetric nuclear matter and beta equilibrium}
\label{sec:beta}

We extend the microscopic results for neutron matter to matter
containing both neutrons and protons. To this end we use for the
energy per particle $\epsilon$ of asymmetric nuclear matter an
expression that interpolates between the properties of symmetric
nuclear matter and neutron matter. For $\epsilon$ we take the kinetic
energy plus an expression for the interaction energy that is quadratic
in the neutron excess $1-2x$:
\begin{eqnarray}
\frac{\epsilon(\bar{n},x)}{T_0} &=& \frac{3}{5} \left[ x^{5/3} + (1-x)^{5/3} \right] (2 \bar{n})^{2/3}
    \nonumber \\[2mm] &&- \left[ ( 2 \alpha - 4 \alpha_L) x (1 - x) + \alpha_L \right] \bar{n}
    + \left[ ( 2 \eta - 4 \eta_L) x (1 - x) + \eta_L \right] \bar{n}^{\gamma} \,, \label{Eskyrme}
\end{eqnarray}
where $\bar{n} = n/n_0$ and $x = n_p/n$ denote the density in units of
the saturation density and the proton fraction, respectively. $T_0 =
(3 \pi^2 n_0/2)^{2/3} \hbar^2/(2m) = 36.84 \, \rm{MeV}$ is the Fermi
energy of symmetric nuclear matter at the saturation density.
Equation~(\ref{Eskyrme}) does not include the contributions from rest masses. 
The corresponding result for the pressure $P=n^2\partial \epsilon/
\partial n$ is
\begin{eqnarray}
\frac{P(\bar{n},x)}{n_0 \, T_0} &=& \frac{2}{5} \left[ x^{5/3} + (1 - x)^{5/3} \right] (2 \bar{n})^{5/3} \nonumber \\[2mm]
    &&- \left[ \left( 2 \alpha - 4 \alpha_L \right) x ( 1- x) + \alpha_L \right] \bar{n}^2
    + \gamma \left[ \left(2 \eta - 4 \eta_L\right) x ( 1- x) + \eta_L \right] \bar{n}^{\gamma+1} \,. \label{Pskyrme}
\end{eqnarray}
The parameters $\alpha, \eta, \alpha_L$ and $\eta_L$ can be determined
from the saturation properties of symmetric nuclear matter combined
with the microscopic calculations for neutron matter of
Section~\ref{sec:neutron_matter}. For $\gamma = 4/3$ and empirical
saturation properties of symmetric nuclear matter,
\begin{equation}
\epsilon(\bar{n}=1, x=1/2) = - B = - 16 \, {\rm MeV} 
\quad {\rm and} \quad 
P(\bar{n}=1, x=1/2) = 0 \,,
\end{equation}
this results in $\alpha = 5.87$, $\eta = 3.81$, and a reasonable
incompressibility parameter
\begin{equation}
K = 9 \left. \frac{\partial^2 \epsilon(\bar{n},x)}{\partial \bar{n}^2} 
\right|_{\bar{n}=1,x=1/2} = 236\,\rm{MeV} \,.
\end{equation}
The precise value of $K$ can be adjusted by modifying the exponent
$\gamma$ in Equations~(\ref{Eskyrme}) and~(\ref{Pskyrme}). However, as
shown in Table~\ref{tab:SvL}, the predicted range for the symmetry
energy and its density derivative depend very weakly on the particular
choice of $\gamma$ and the resulting value of $K$, because the leading
density dependence of the symmetry energy is linear.

\begin{figure}[t!]
\begin{center}
\includegraphics[scale=0.8,clip=]{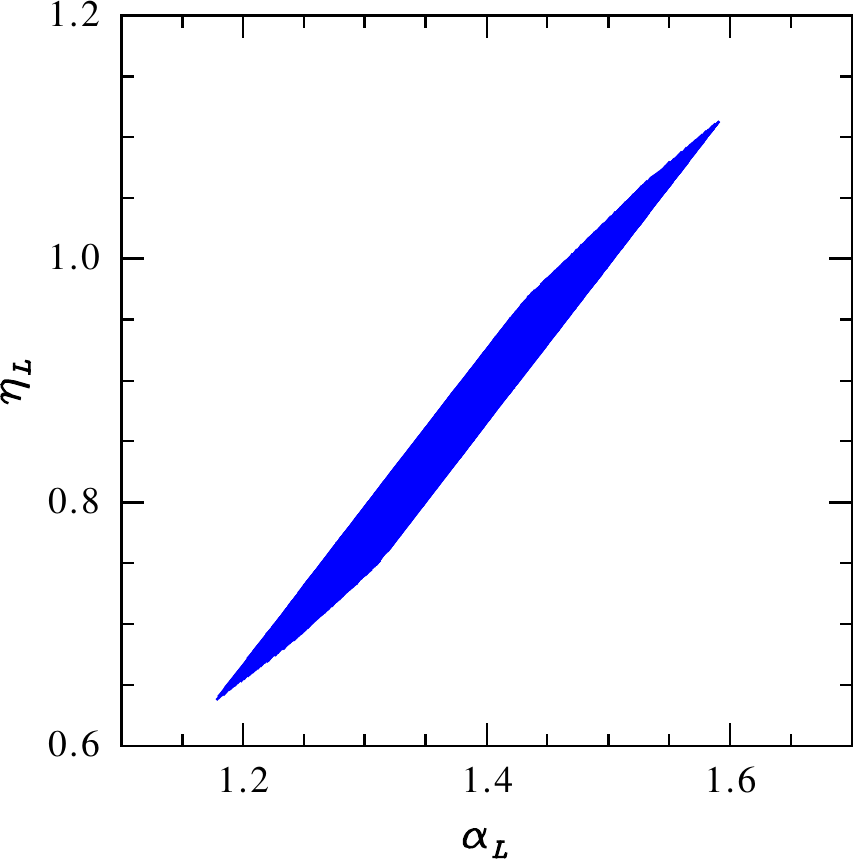}
\end{center}
\caption{(Color online) Allowed range for $\alpha_L$ and $\eta_L$
of the parametrizations~(\ref{Eskyrme}) and~(\ref{Pskyrme})
fit to the saturation point of symmetric nuclear matter and to the
calculated neutron matter energy and pressure.\label{alpha_eta}}
\end{figure}

\begin{table}[t!]
\begin{center}
\begin{tabular}{cc|c|c}
$\gamma$ & $K$ [MeV] & $S_v$ [MeV] & $L$ [MeV] \\
\hline
$1.2$ & 210 & $29.7 - 32.8$ & $32.4 - 53.4$ \\
$4/3$ & 236 & $29.7 - 33.2$ & $32.5 - 57.0$ \\
$1.45$ & 260 & $30.1 - 33.5$ & $33.6 - 56.7$
\end{tabular}
\end{center}
\caption{Predicted range for the symmetry energy $S_v$ and for the
$L$ parameter, which determines the density dependence of the 
symmetry energy. Results are given for different $\gamma$ values, 
which lead to different incompressibilities $K$, but, as shown,
the predicted ranges for $S_v$ and $L$ depend very weakly on $\gamma$.
\label{tab:SvL}}
\end{table}

The parameters $\alpha_L$ and $\eta_L$ are extracted from the
calculated bands for the neutron matter energy and pressure of
Figures~\ref{EN_nobeta} and~\ref{p_nobeta}. Our results are based on
the blue bands (with renormalization-group evolution), unless stated
otherwise. We have first checked that the
parametrizations~(\ref{Eskyrme}) and~(\ref{Pskyrme}) provide excellent
global fits for the energy and pressure up to a density $n_1 \approx
1.1 \, n_0$. To determine $\alpha_L$ and $\eta_L$, we sample their
values systematically and require that the resulting energy and
pressure be within the uncertainty bands shown in
Figures~\ref{EN_nobeta} and~\ref{p_nobeta} for densities from $0.45 \,
n_0$ to $1.1 \, n_0$.  This leads to the allowed range for $\alpha_L$
and $\eta_L$ shown in Figure~\ref{alpha_eta}, with correlated limits
$\alpha_L = 1.18 - 1.59$ and $\eta_L = 0.64 - 1.11$.

The proton fraction $x$ for matter in beta equilibrium is determined
by minimizing, for a given nucleon density, the total energy per
particle, Equation~(\ref{Eskyrme}), plus the contributions from 
electrons and from the rest mass of the nucleons. 
This amounts to the condition that $\mu_n + m_n c^2 = \mu_p
+ m_p c^2 + \mu_e$, where $\mu_n$ and $\mu_p$ are the neutron and
proton chemical potentials without the rest mass contribution, or
equivalently
\begin{equation}
\frac{\partial \epsilon(\bar{n},x)}{\partial x} + \mu_e(\bar{n},x) 
- (m_n - m_p) c^2 = 0 \,.
\end{equation}
For an ultrarelativistic, degenerate electron gas, the chemical
potential is given by $\mu_e(\bar{n},x) = \hbar c \, (3 \pi^2 x n_0
\bar{n})^{1/3}$. The allowed ranges for $\alpha_L$ and $\eta_L$ imply
ranges for the proton fraction and the neutron and proton chemical
potentials in beta equilibrium, which are given for the saturation
density $n_0$ and for $n_0/2$ in Table~\ref{tab:pfrac}. In the
calculations we neglected the difference between the neutron and proton
masses (1.3~MeV), which is small compared with $\mu_e \sim 100 \, {\rm
MeV}$. These ranges provide anchor points for other equations of state.

\begin{table}[t]
\begin{center}
\begin{tabular}{c|c|c|c}
$n=n_0$ & $x$ & $\mu_n$ [MeV] & $\mu_p$ [MeV] \\
\hline
min & $0.040$ & $54.2$ & $-58.0$ \\
max & $0.053$ & $51.9$ & $-71.5$ \\
\hline
$n=n_0/2$ \\
\hline
min & $0.030$ & $34.6$ & $-46.1$ \\
max & $0.033$ & $34.3$ & $-48.7$ 
\end{tabular}
\end{center}
\caption{Proton fraction $x$ and chemical potentials $\mu_n$ and $\mu_p$ 
in beta equilibrium for the saturation density $n_0$ and for $n_0/2$. 
The rows marked ``min'' and ``max'' give the range of the uncertainty
band.\label{tab:pfrac}}
\end{table}

\begin{figure}[t]
\begin{center}
\includegraphics[scale=0.45,clip=]{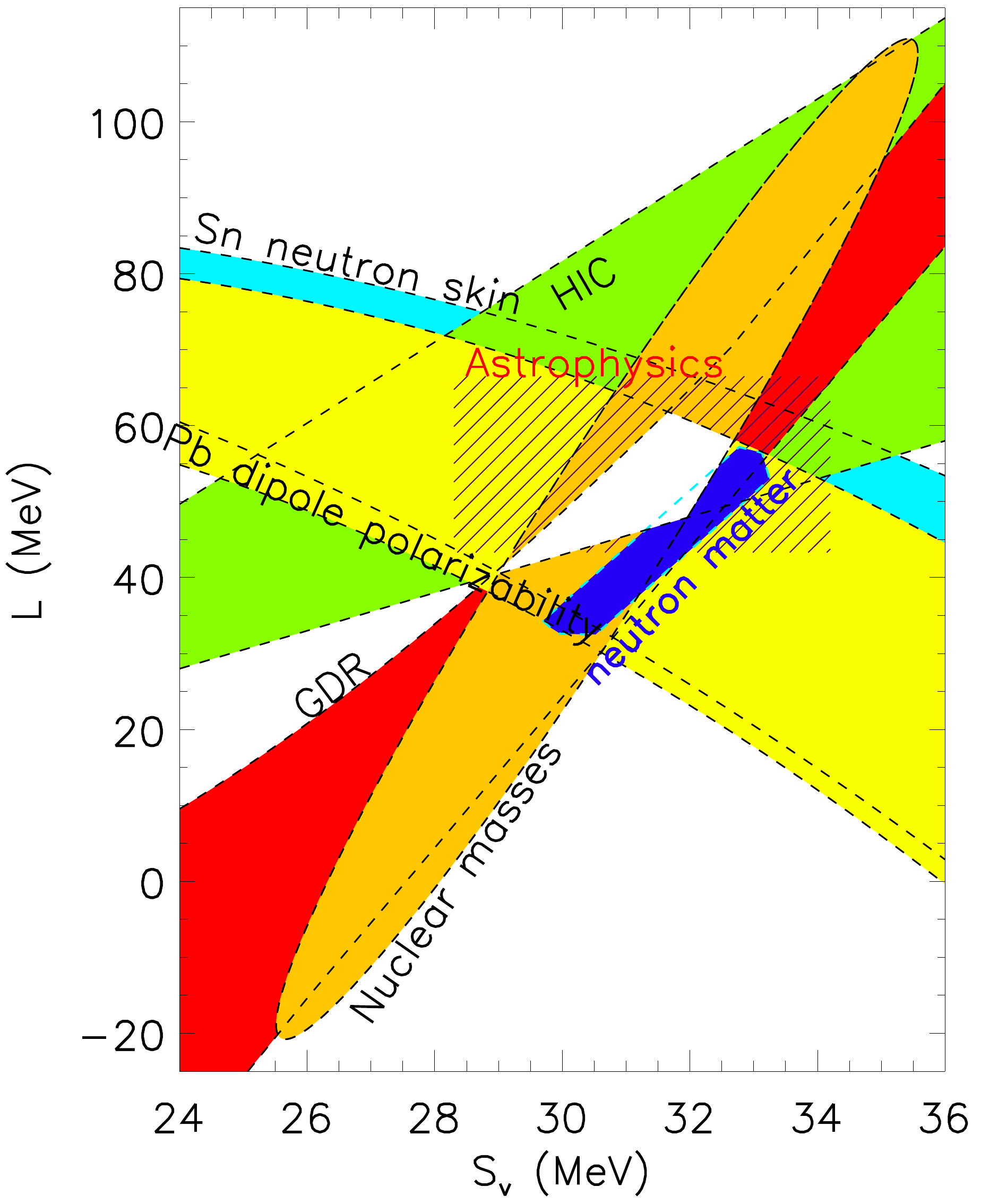}
\end{center}
\caption{(Color online) Constraints for the symmetry energy $S_v$ and
the $L$ parameter following \citet{LL}. The blue region shows our 
neutron matter constraints, in comparison to bands based on different
empirical extractions (for details see text). The white area gives
the overlap region of the different empirical ranges.\label{symcor}}
\end{figure}

The parametrizations~(\ref{Eskyrme}) and~(\ref{Pskyrme}) also make it
possible to reliably extract the symmetry energy $S_v$ and its
density derivative $L$,
\begin{equation}
S_v = \frac{1}{8} \frac{\partial^2 \epsilon(\bar{n},x)}{\partial x^2} 
\biggr|_{\bar{n}=1, x=1/2} \quad {\rm and} \quad 
L = \frac{3}{8} \frac{\partial^3 \epsilon(\bar{n},x)}{\partial \bar{n} 
\partial x^2} \biggr|_{\bar{n}=1, x=1/2} \,.
\end{equation}
The region for $\alpha_L$ and $\eta_L$ translates into an allowed
region for the symmetry energy $S_v$ and the $L$ parameter shown in
Figure~\ref{symcor}, after \citet{LL}. In addition, we give in
Table~\ref{tab:SvL} the predicted ranges for $S_v$ and $L$ for
different $\gamma$ values, %(Figure~\ref{symcor} shows the results for
%$\gamma=4/3$), 
corresponding to different incompressibilities $K=210 \mev$, $236
\mev$ and $260 \mev$. The predicted range for $\gamma = 4/3$ nearly
spans the ranges for the other $\gamma$ values. This demonstrates that
the extrapolation~(\ref{Eskyrme}) is robust and that the theoretical
uncertainty due to the choice of $\gamma$ is very weak and clearly
much smaller than the empirical bands shown in Figure~\ref{symcor}.

In Figure~\ref{symcor}, we compare the $S_v$ and $L$ region predicted
by our neutron matter results with values extracted from other
data~\citep{LL}. It is striking that the neutron matter results lead
to the strongest constraints. These agree well with constraints
obtained from energy-density functionals for nuclear masses (orange
band)~\citep{masses} and from the $^{208}$Pb dipole polarizability
(yellow band)~\citep{Tamii}. In addition, there is good agreement with
studies of the Sn neutron skin (light blue band)~\citep{Sn}, of
isotope diffusion in heavy ion collisions (HIC, green
band)~\citep{HIC}, and of giant dipole resonances (GDR, red
band)~\citep{GDR,LL}. Moreover, there is very good agreement with an
estimate obtained from modeling X-ray bursts and quiescent low-mass
X-ray binaries (shaded region, labeled
`Astrophysics')~\citep{Steiner1}. Remarkably, the constraints from
these analyses have a common area of intersection (white area), which
overlaps within uncertainties with the constraints from microscopic
calculations of neutron matter. This suggests that quartic and higher-order 
corrections are relatively small. In the future we shall study them on 
the basis of chiral EFT. Based on the comparison in Figure~\ref{p_nobeta}, there 
is also very good agreement with the $S_v- L$ correlation band obtained from 
the Quantum Monte Carlo results of~\citet{Gandolfi} [see~\citet{LL}].

\section{Crust-core boundary}
\label{sec:crust-core}

The transition between the neutron star crust and a uniform state in
the core is a first-order phase transition. However, the density jump
across the boundary is small; consequently, a good estimate for the
density at which the transition from crustal matter to uniform matter
in the core takes place may be obtained by determining the conditions
under which, on lowering the density, matter becomes unstable to
formation of a small density modulation \citep{BBP,Chris_instability}.
The density at which this occurs provides a lower bound on the density
of the uniform phase at which the transition occurs.

The electron screening length in the crust is large compared with
typical nuclear separations, so it is a good first approximation to
regard the electron density as remaining uniform when a proton density
wave is formed. For matter to be stable to formation of a long
wavelength density fluctuation, the energy density must increase when
the density modulations are imposed subject to the condition that the
total number of neutrons and the total number of protons remain
constant. Our discussion is adapted from \citet{BBP} with minor
variations. One condition for stability when Coulomb and contributions
to the energy from density gradients are included is
\begin{equation}
v_0 + 2 (4\pi e^2\beta)^{1/2} - \beta k_{\rm FT}^2 > 0 \,.
\label{BBP}
\end{equation}
The quantity
\begin{equation}
v_0 = \frac{\partial\mu_p}{\partial n_p} - \frac{(\partial\mu_p/
\partial n_n)^2}{(\partial\mu_n/\partial n_n)}
\end{equation}
is an effective proton-proton interaction when the Coulomb interaction
is neglected. The first term is the contribution when the neutron
density is constant and the second term is an induced interaction due
to exchange of neutron density fluctuations. The effect of
inhomogeneities in the density distribution is described by the
quantity $\beta$, given by
\begin{equation}
\beta=2 (Q_{pp}+2Q_{np}\zeta+Q_{nn}\zeta^2) \,,
\end{equation}
and a more detailed discussion is given in Appendix~\ref{instability}.
Here $Q_{ij}=B_{ij}/n_0$ in the notation of \citet{BBP} and
\begin{equation}
\zeta=-\frac{\partial\mu_p/\partial n_n}{\partial\mu_n/\partial n_n} \,.
\end{equation}
The Thomas-Fermi wave number $k_{\rm TF}$ is given by
\begin{equation}
k_{\rm TF}^2 = \frac{4}{\pi} \frac{e^2}{\hbar c} \, k_e^2 \,,
\end{equation}
with $k_e = (3 \pi^2 n x)^{1/3}$. Generally, the second term in
Equation~(\ref{BBP}), which is due to the nonuniformity of the electron
density, is small compared with the first but we do not drop it. If
Equation~(\ref{BBP}) and the condition $\partial \mu_n/\partial n_n
>0$ are satisfied, matter is stable to small density modulations. With
decreasing density, for realistic equations of state, the first
stability condition to be violated is Equation~(\ref{BBP}), and
instability first sets in when this becomes an equality, which
corresponds to Equation~(9.18) of \citet{BBP}.

\begin{figure}[t]
\begin{center}
\includegraphics[scale=0.375,clip=]{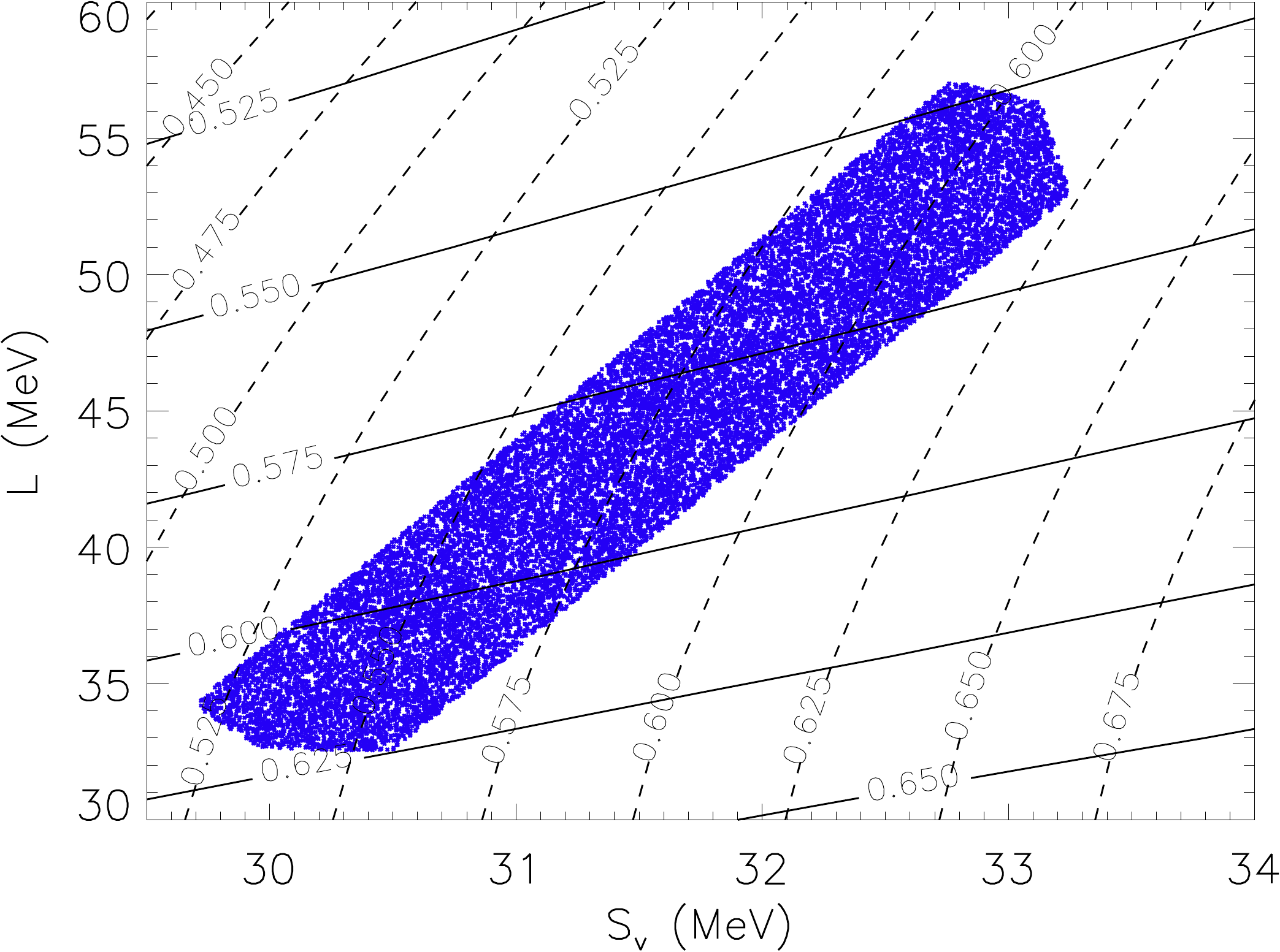}
\hspace*{2mm}
\includegraphics[scale=0.375,clip=]{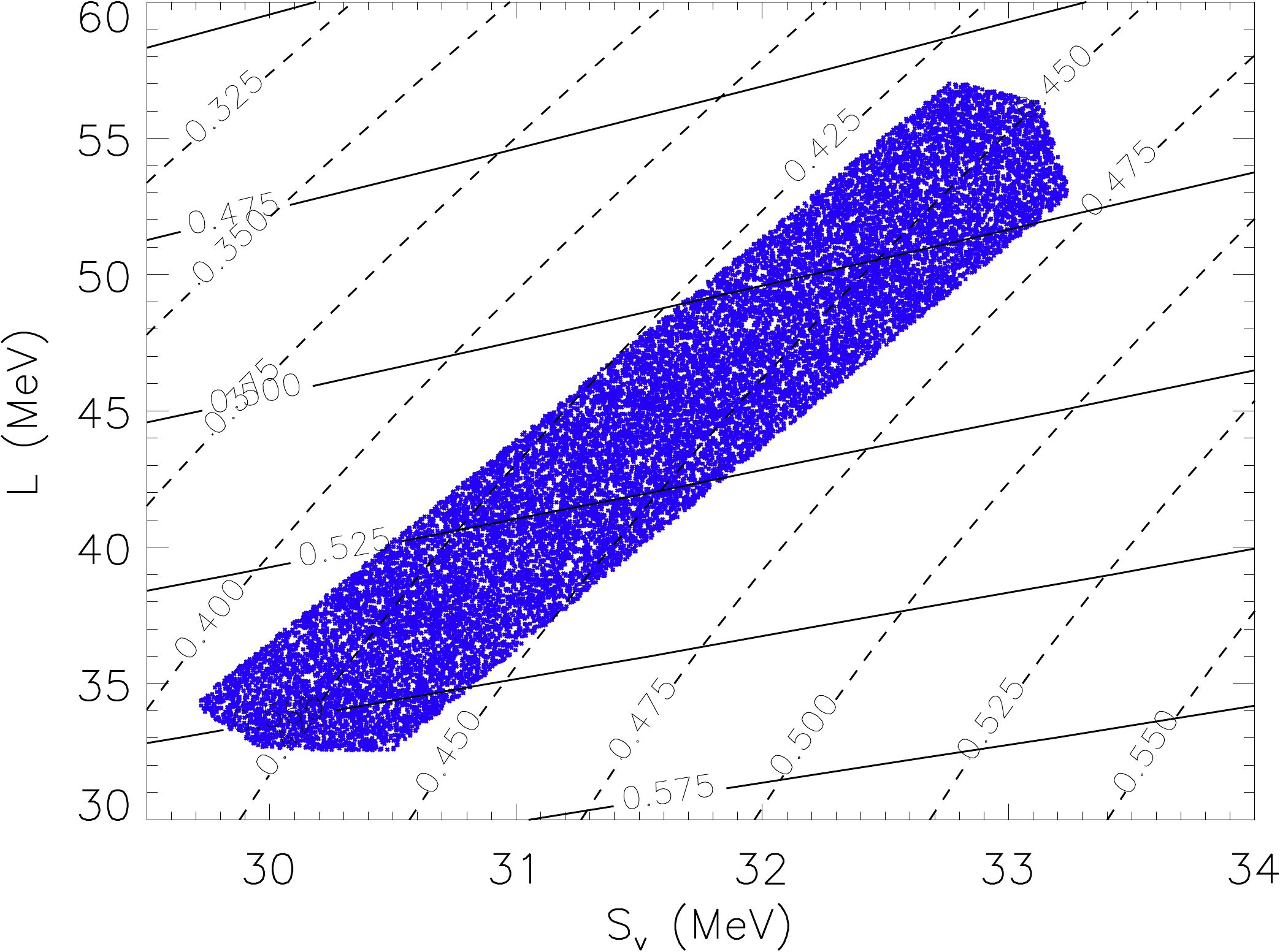}
\end{center}
\caption{(Color online) Contours of the lower bounds on the density
of the liquid phase (solid lines) and the pressure (dashed lines) at
the crust-core boundary in the $S_v-L$ plane. The density is measured
in terms of the saturation density $n_0$ and the pressure in 
${\rm MeV} \fmiq$. In the left panel, Coulomb and density gradient
terms are neglected ($Q=0$) and in the right panel $Q=75 \mev \, 
{\rm fm}^5$. The blue areas are the allowed region in Figure~\ref{symcor}.
\label{transsl}}
\end{figure}

Figure~\ref{transsl} shows contour plots of the densities and
pressures at which instability sets in for uniform matter in beta
equilibrium, with Coulomb and density gradient contributions for $Q=75
\mev \, {\rm fm}^5$ (right panel), and without these contributions,
$Q=0$ (left panel), see Appendix~\ref{instability}. We find transition
densities around $\bar{n} = 0.55-0.625$, and inclusion of the Coulomb
and density gradient contributions lowers these by about $15\%$ to
$\bar{n} = 0.475-0.55$.  These transition densities are somewhat
smaller than those predicted by the FPS and FPS21 interactions in
\citet{Chris_instability}, and consequently neutron star models with
the interactions used in this paper will have lower crustal masses and
lower crustal moments of inertia.

\section{General extension}
\label{sec:gen_ext}

\begin{figure}[t] 
\begin{center}
\includegraphics[scale=0.9,clip]{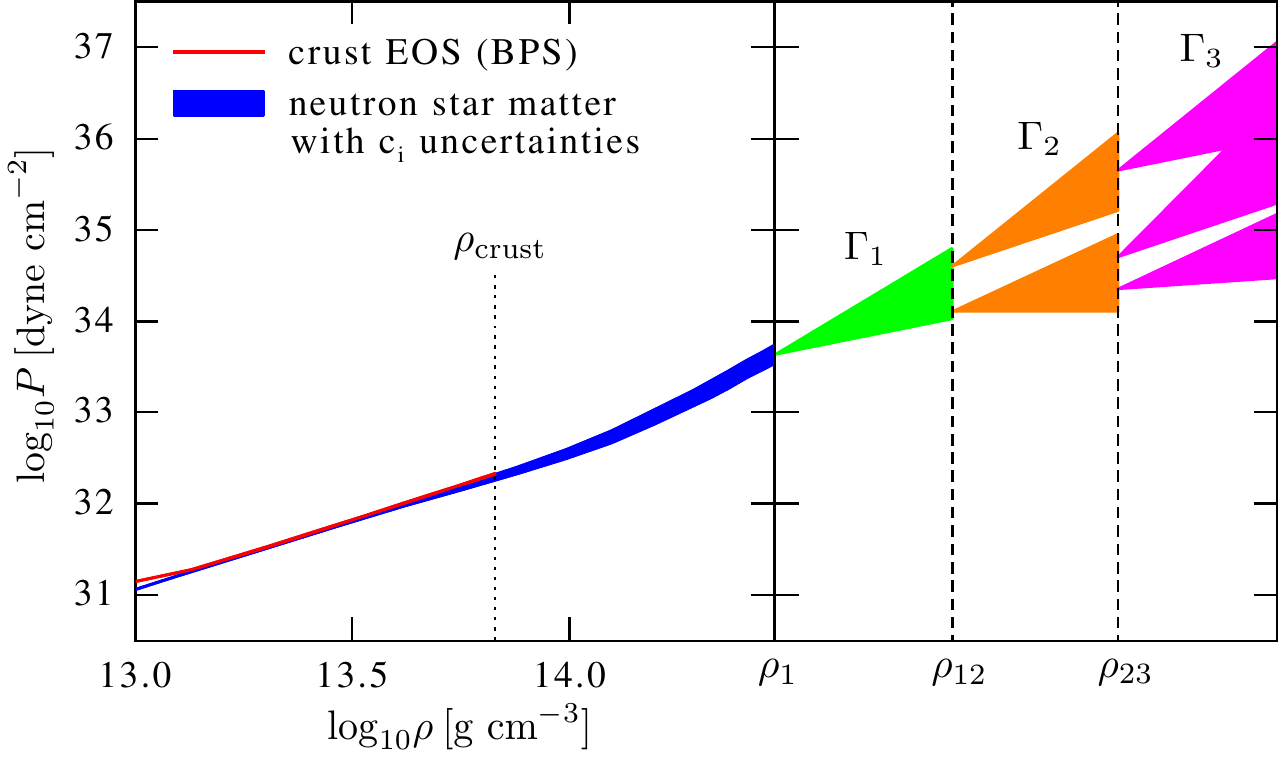}
\end{center}
\caption{(Color online) Pressure $P$ of neutron star matter as a
function of mass density~$\rho$. The left part compares the pressure
band predicted from the neutron matter results of
Figure~\ref{p_nobeta} and incorporating beta equilibrium following
Section~\ref{sec:beta} to the BPS outer crust EOS~\citep{BPS,Vautherin}.
The right part illustrates the general piecewise polytropic extension
of the neutron star EOS to higher densities. We take $\rho_{\rm crust}
= \rho_0/2$, $\rho_1 = 1.1 \, \rho_0$, and vary the polytrope
parameters over the ranges $1 \leqslant \Gamma_1 \leqslant 4.5$,
$1.5 \, \rho_0 \leqslant \rho_{12} \leqslant 8 \, \rho_0$, $0 \leqslant
\Gamma_2 \leqslant 8$, $\rho_{12} \leqslant \rho_{23} \leqslant
8.5 \, \rho_0$, and $0.5 \leqslant \Gamma_{23} \leqslant 8$ (see
text for details).\label{fig:polytropic}}
\end{figure}

We investigate the structure of nonrotating neutron stars by solving
the Tolman-Oppen-heimer-Volkov equations. Since the central densities
of neutron stars can significantly exceed the regime of our neutron
matter calculations, we need to extend the 
EOS beyond this density. To this end, we employ a general polytopic
extension, where the pressure of neutron star matter is piecewise
given by $P(\rho) = \kappa\,\rho^{\Gamma}$, with mass density $\rho
=mn$~\citep{poly,Kai}. This is illustrated in Figure~\ref{fig:polytropic}.
At low densities we use the BPS crust EOS~\citep{BPS,Vautherin} up to
the transition density $\rho_{\rm crust}$. Figure~\ref{fig:polytropic}
shows that for densities $\rho_0/10$ to $\rho_0/2$ the BPS crust EOS
lies within the band predicted for the pressure of neutron star matter
based on the neutron matter results and incorporating beta equilibrium
following Section~\ref{sec:beta} [see also \citet{Kai}].
Therefore our results are insensitive to the particular choice of
$\rho_{\rm crust}$ within this region. In the following, we use
$\rho_{\rm crust} = \rho_0/2$, based on our results for the crust-core
boundary given by Figure~\ref{transsl}. The pressure from $\rho_{\rm
crust}$ to $\rho_1=1.1 \, \rho_0$ is given by the band predicted by
chiral EFT interactions. Our results are insensitive to the precise
value of $\rho_1$ in the vicinity of saturation density $\rho_0$, so
we have taken a conservative value for which the uncertainty band of
the microscopic neutron matter calculations is reasonable.

For the extension beyond $\rho_1$, we use three polytropes with
exponents $\Gamma_1, \Gamma_2$ and $\Gamma_3$, which make it possible
to vary the soft- or stiffness of the EOS in the density regions
1:~$\rho_1 \leqslant \rho \leqslant \rho_{12}$; 2: $\rho_{12}
\leqslant \rho \leqslant \rho_{23}$, and 3: $\rho \geqslant
\rho_{23}$, respectively (see Figure~\ref{fig:polytropic}). For
densities just above $\rho_1$, the EOS is still rather well
constrained by the neutron matter calculations. The band predicted for
the pressure of neutron star matter at $\rho_1$ corresponds to values 
of $\Gamma$ in the range $2.25 - 2.5$. Therefore, we take a restricted
range for the first polytropic exponent $1.0 \leqslant \Gamma_1
\leqslant 4.5$. We vary the value of all $\Gamma_i$ in steps of
$0.5$. At intermediate densities we allow for the possibility of a
phase transition and take a broad range $0 \leqslant \Gamma_2
\leqslant 8$. Finally, for densities beyond $\rho_{23}$ we allow for
$0.5 \leqslant \Gamma_3 \leqslant 8$.  We exclude the value 0 for this
density region in order to avoid artefacts connected with a phase
transition up to arbitrarily high density. For the densities between
polytropes, $\rho_{12}$ and $\rho_{23}$, we take $1.5 \, \rho_0
\leqslant \rho_{12} < \rho_{23} < \rho_{\rm max}$ with a step size of
$\rho_0/2$. We will show in the next section that the maximal density
for our suite of EOSs of neutron stars is $\rho_{\rm max} \approx 8.3
\, \rho_0$.

The general polytropic extension leads to a very large number of EOSs,
which cover all possible pressures in the grey region
in Figure~\ref{prho_extension}. Furthermore we emphasize that
this strategy is very general. It is based on a well defined
uncertainty band at nuclear densities and does not rely on assumptions
about the nature of the constituents of neutron star matter and their
interactions at higher densities. The values of $\Gamma_i$ and
$\rho_{ij}$ are limited by nuclear physics and observation. In the
following, we will demonstrate that the recent observation of a $1.97
\pm 0.04 \, M_{\odot}$ neutron star~\citep{Demorest} puts rather tight
constraints on the EOS at high densities and the radii of neutron
stars. We note that our results agree with those of our first study
\citep{Kai}, which used only two polytropes. This shows that the
general extension is robust and the conclusions would not change
significantly if additional polytropes were introduced to characterize
the pressure at high densities.

\section{Constraints on the nuclear equation of state and neutron stars}
\label{sec:constraints}

\begin{figure}[t]
\begin{center}
\includegraphics[scale=0.725,clip=]{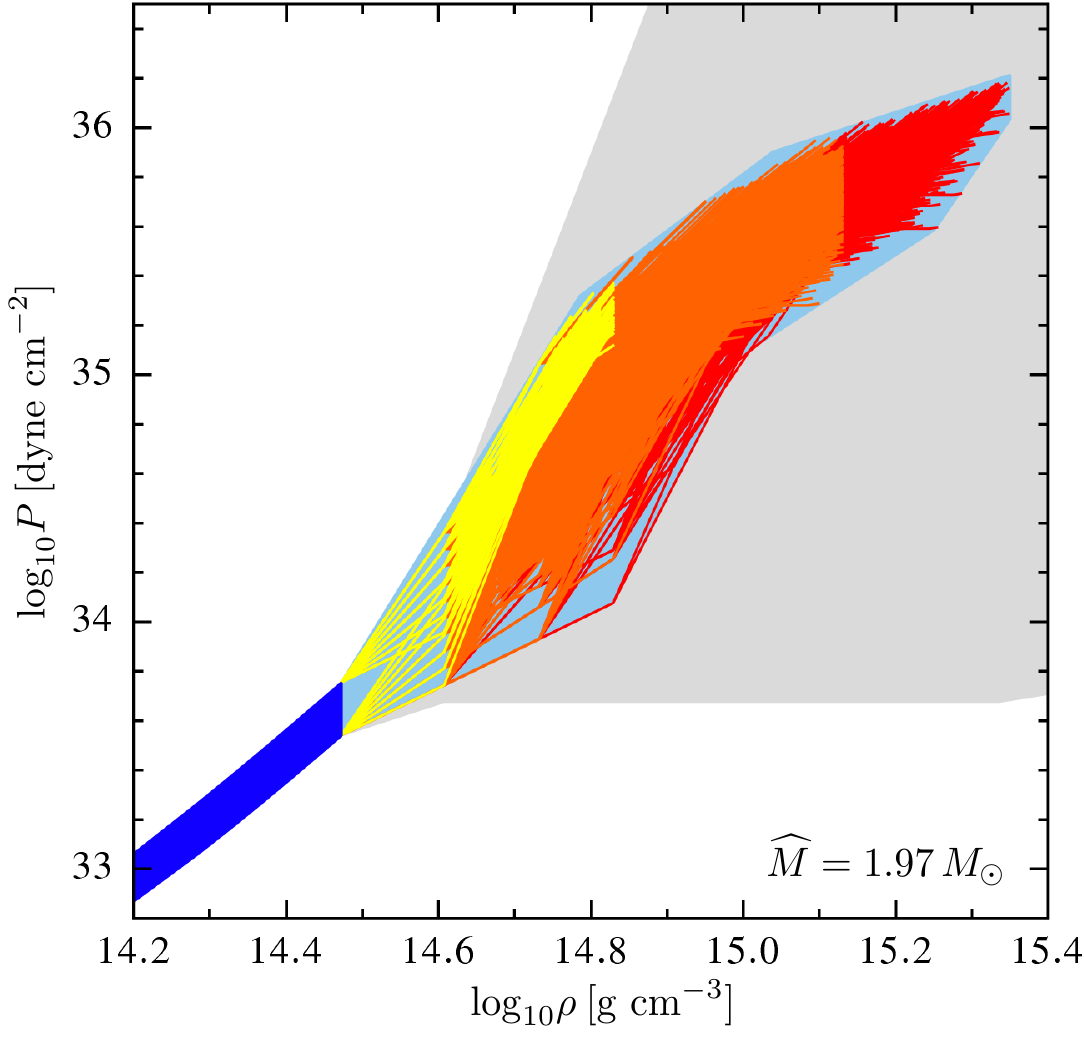}
\hspace{2mm}
\includegraphics[scale=0.725,clip=]{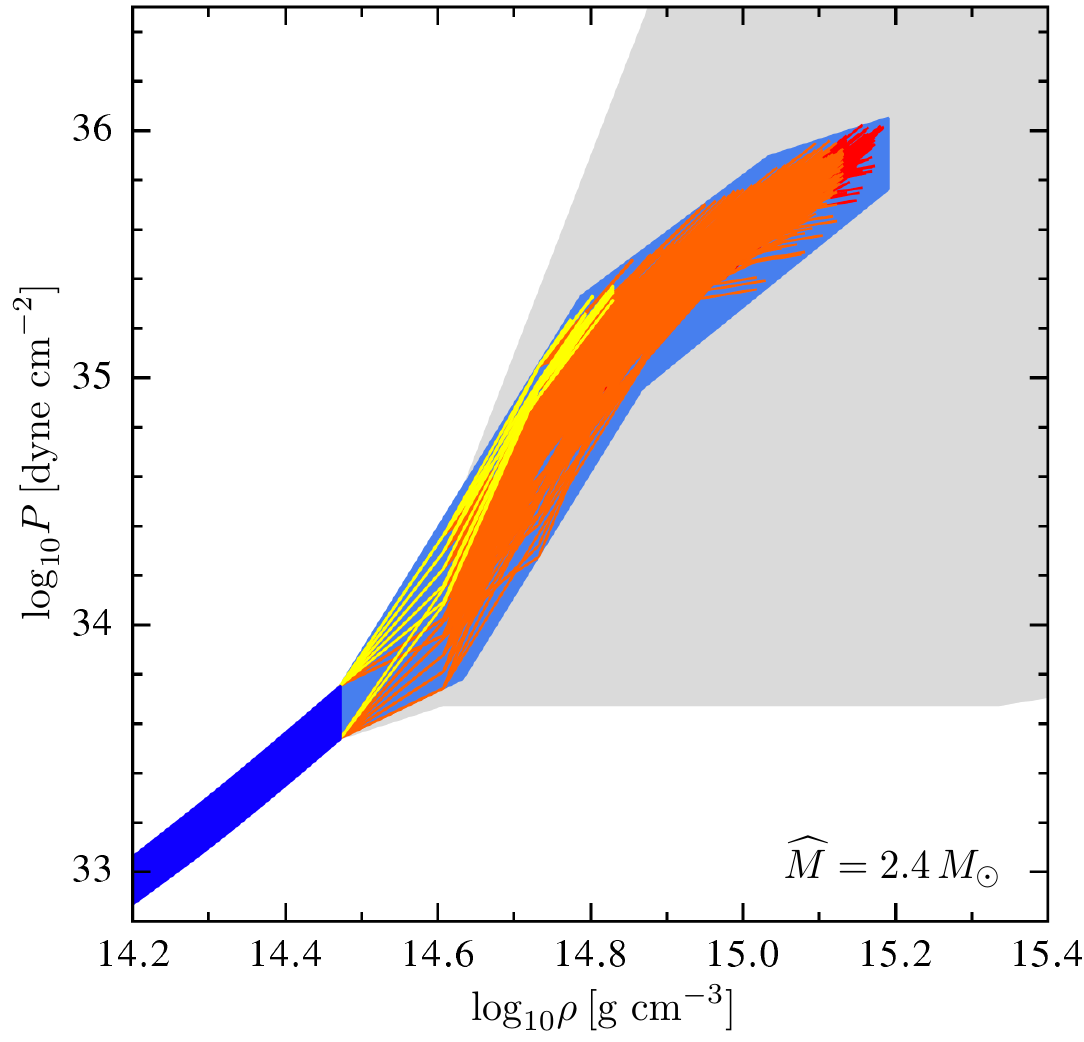}
\end{center}
\caption{(Color online) Pressure $P$ of neutron star matter as a
function of the mass density~$\rho$. The blue band at lower densities
represents the pressure predicted by the neutron matter results
of Figure~\ref{p_nobeta} with beta equilibrium incorporated as described 
in Section~\ref{sec:beta}. The grey area shows the pressure accessed by
the possible polytropic extensions. The lighter 
blue band is the envelope of the general 
polytropic extensions that are causal and support a neutron
star of mass $\widehat{M} = 1.97 \, M_\odot$ (left panel) and $\widehat{M} =
2.4 \, M_\odot$ (right panel). We also show the individual EOSs within
these bands: The lines denote EOSs with central densities
$\rho_c \leqslant 2.5 \, \rho_0$ (yellow), for $2.5 \, \rho_0 < 
\rho_c \leqslant 5 \, \rho_0$ (orange), and for $\rho_c > 
5 \, \rho_0$ (red).\label{prho_extension}}
\end{figure}

The piecewise polytropic extension described in the previous section
is used to generate a very large number of equations of state that
cover the pressure-density plane at higher densities. We solve the
Tolman-Oppenheimer-Volkov equations for each of these EOSs and retain
only those that fulfill the following two constraints:
\begin{itemize}
\item[1.)] the speed of sound remains smaller than the speed of light
for all densities relevant in neutron stars: $v_s(\rho) =
\sqrt{dP/d{\mathcal E}} \, c \leqslant c$, where ${\mathcal E}$ is the 
energy density.
\item[2.)] the EOS supports a neutron star mass $M = \widehat{M}$, 
the mass of the heaviest neutron star observed.
\end{itemize}

We consider each EOS up to densities at which the maximal neutron star
mass is reached or the EOS becomes acausal, whichever density is
smaller. In Figure~\ref{prho_extension} we present the individual EOSs
that fulfill both constraints for two cases: $\widehat{M} = 1.97 \,
M_\odot$ (left panel), the mass of the heaviest known neutron
star~\citep{Demorest}, and $\widehat{M} = 2.4 \, M_\odot$ (right
panel), an estimated mass of the black widow pulsar
B1957+20~\citep{Kerkwijk}. However, since the uncertainties of the
latter determination are large, the $\widehat{M} = 2.4 \, M_\odot$
constraint should be considered as a hypothetical mass, which is used
here to probe the sensitivity of our results to the constraint from
observations.

\begin{figure}[t]
\begin{center}
\includegraphics[scale=0.8,clip=]{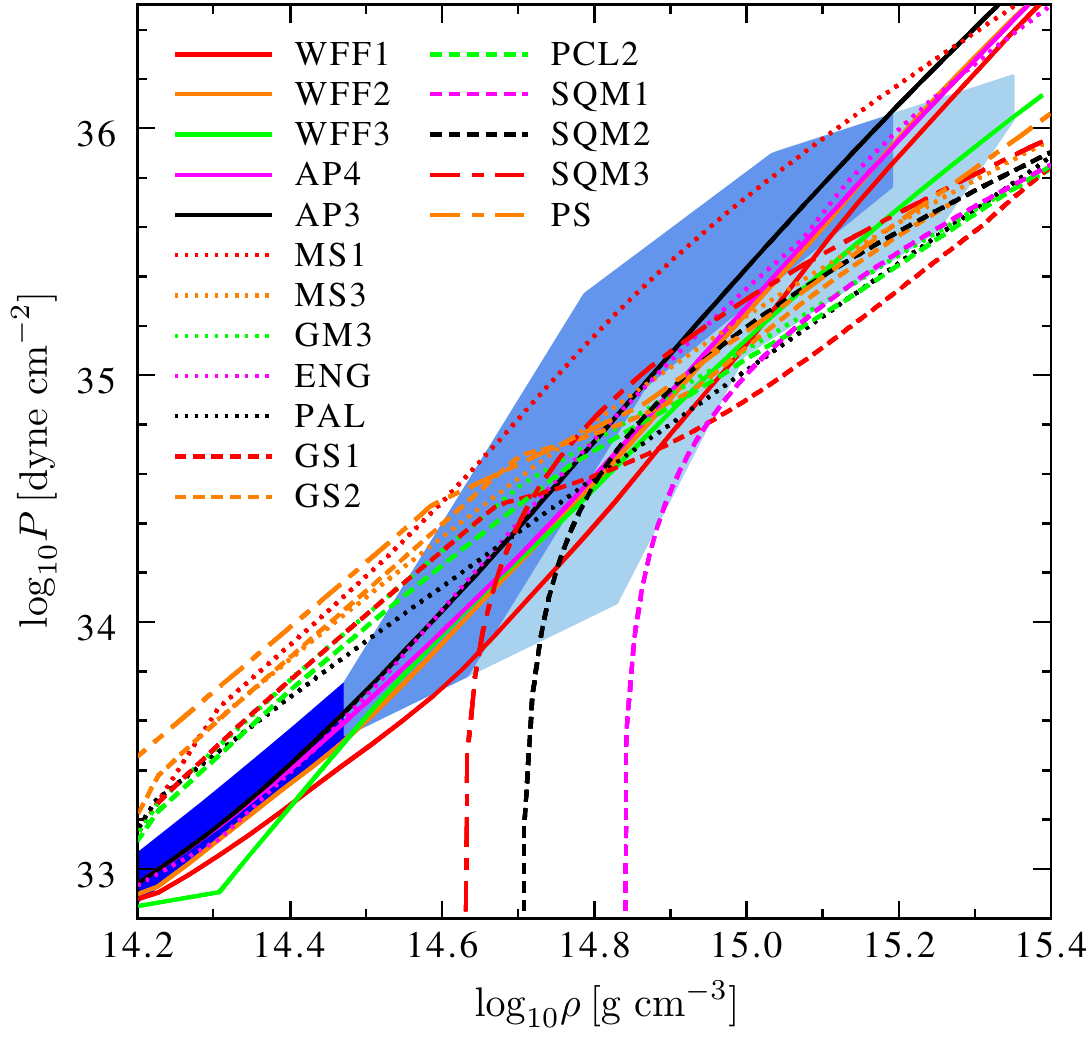}
\end{center}
\caption{(Color online) Constraints for the pressure $P$ of neutron 
star matter as a function of mass density $\rho$ compared to EOSs
commonly used to model neutron stars~\citep{LP}. The color coding of
the bands is as in Figure~\ref{prho_extension}.\label{prho_compare}}
\end{figure}

The blue band at lower densities in Figure~\ref{prho_extension}
represents the pressure predicted for matter in beta equilibrium as
described in Sections~\ref{sec:neutron_matter} and
~\ref{sec:beta}. The lighter blue bands give the allowed EOS range,
which is the envelope of the allowed polytropes at higher
densities. The color of the individual lines indicates the maximal
central density of the individual EOS~(see Figure caption). Clearly,
the pressure accessed by the possible polytropic extensions (the grey
area in Figure~\ref{prho_extension}) is substantially reduced by
causality and by the constraint from neutron star mass
measurements. The higher the mass of the heaviest neutron star
observed, the stronger the EOS band is contrained. In addition, we
find that the maximal densities in neutron stars are limited: stiff
EOSs with large polytropic exponents have smaller maximal
densities~(see yellow lines), which are strongly constrained by
causality.  Softer EOSs tend to have larger central densities. For $M
= 1.97 \, M_{\odot}$ we find central densities as high as
$\approx 8.3 \, \rho_0$, and, for $M = 2.4 \, M_{\odot}$ the
densities reach only $\approx 5.8 \, \rho_0$. 
%These values are to be
%compared with the upper limits $\approx 9.7 \, \rho_0$ and $\approx
%6.5 \, \rho_0$, respectively, derived purely from causality arguments
%in \cite{LP2011}.

In Figure~\ref{prho_compare} we compare the EOS bands of
Figure~\ref{prho_extension} with a representative set of EOSs used in
the literature. This set contains EOSs calculated within different
theoretical approaches and based on different degrees of freedom. For
details and notation we refer the reader to~\citet{LP}. We find that a
significant number of EOSs are not compatible with the lower density
band based on chiral EFT interactions. In addition, at higher
densities only very few EOSs, including the variational EOSs based on
phenomenological nuclear potentials~\citep{APR}, AP3 and AP4 in
Figure~\ref{prho_compare}, are located within the uncertainty bands
over the entire density range.

\begin{figure}[t]
\begin{center}
\includegraphics[scale=0.8,clip=]{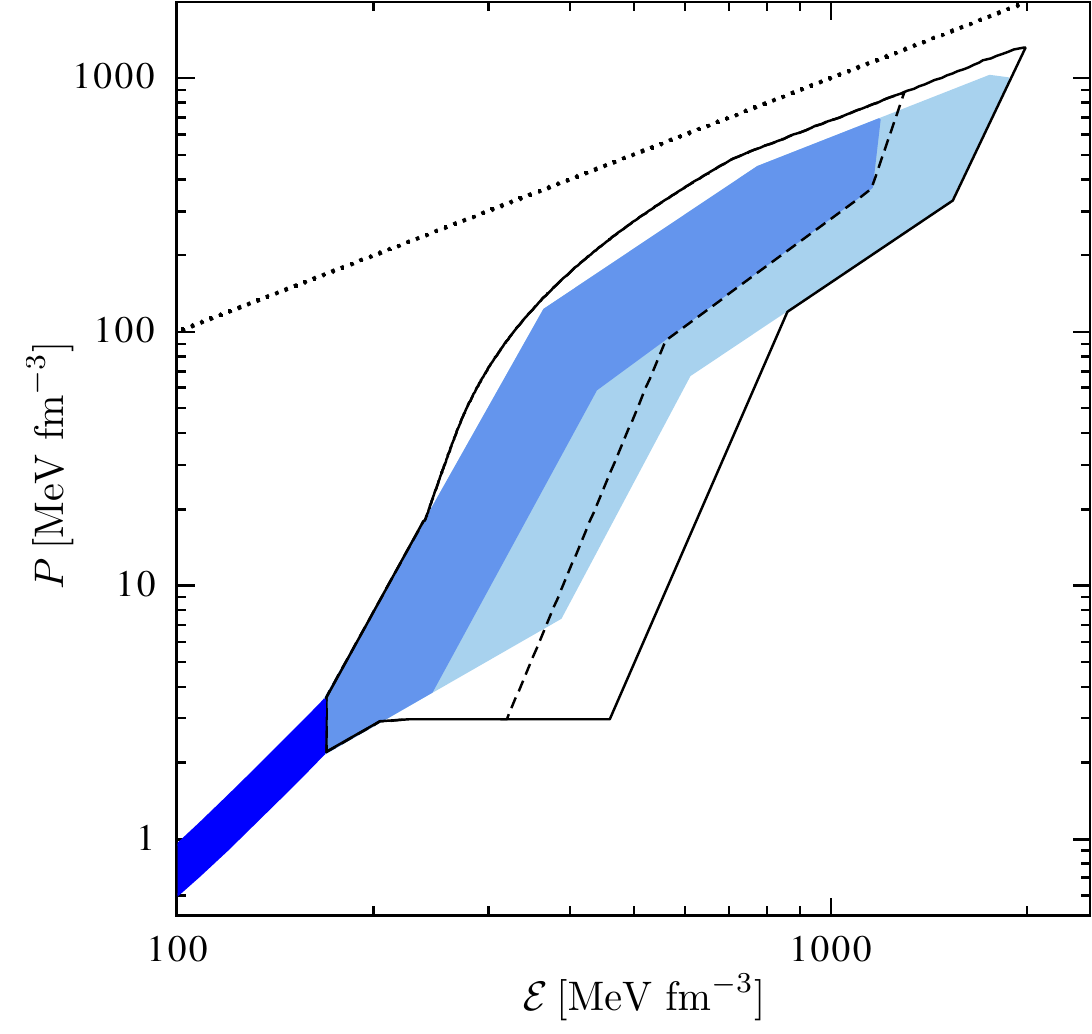}
\end{center}
\caption{(Color online) Constraints on the pressure $P$ of neutron 
star matter as a function of energy density ${\mathcal E}$. The color
coding of the bands is as in Figure~\ref{prho_extension}. The solid and
dashed lines mark the allowed EOS range using the causal extension
(see text) for $\widehat{M} = 1.97 \, M_{\odot}$ (solid lines) 
and $\widehat{M} = 2.4 \, M_{\odot}$ (dashed lines). The dotted line gives
the causal limit $P = {\mathcal E}$.\label{pepsilon}}
\end{figure}

Figure~\ref{pepsilon} shows the uncertainty bands for the pressure as
a function of energy density ${\mathcal E}$. These are the natural
variables to study to what extent the causality constraint is
responsible for the apparent softening of the EOSs at high
densities. For comparison we show the limit $P = {\mathcal E}$, 
represented by the dotted line. Furthermore, we generated a causal 
extension for each individual EOS for energy densities 
$\mathcal{E} > \mathcal{E}_{\rm limit}$ by choosing 
$P(\mathcal{E})=\mathcal{E}-\mathcal{E}_{\rm limit}+P(\mathcal{E}_{\rm limit})$. 
This ensures that the energy density, pressure, and speed of sound 
are continuous at all densities, with speed of sound $v_s({\mathcal E}) = c$ 
for ${\mathcal E} > {\mathcal E}_{\rm limit}$. This can generate an EOS that has 
the speed of sound equal to the speed of light immediately after a
phase transition at the start of the third polytrope. We exclude such a scenario.
In this way we can extend all EOSs to arbitrarily high densities and probe the
role of the causality constraint on the results. In
Figure~\ref{pepsilon}, the solid lines mark the allowed EOS range
using the causal extension for $\widehat{M} = 1.97 \, M_{\odot}$, the
dashed lines the range for $\widehat{M} = 2.4 \, M_{\odot}$. The
comparison to the blue bands shows that the causal extension changes
only slightly the upper pressure limit and leads to somewhat higher
maximal possible densities in a neutron star. 
%In addition, for
%asymptotically high energy densities, the uncertainty bands approach
%by construction the causal limit $P = {\mathcal E}$ at the upper limit
%of the pressure range. 
We also observe that the causal extensions have
a stronger impact on the lower limit of the pressure uncertainty
band. For EOSs in this region the speed of sound reaches the speed of
light already for small neutron star masses. By employing causal
extensions, the maximal neutron star mass for such EOSs can increase
significantly and consequently more EOSs fulfill the mass constraint,
which explains the significant difference between the blue bands and
the regions within the black lines.
%Along the causal extension we 
%have no information about the baryon density and therefore it is not possible 
%to describe the equation of state there in terms of a (density-dependent) 
%polytropic index $d\ln P/d\ln n$.
%Along the causal extension the
%polytropic index is in general changing with density, ensuring that
%$v_s(\rho) = c$ in this density region. Such a scenario is artificial
%and it is an open question if such EOSs are of 
%practical significance.

\begin{figure}[t!]
\includegraphics[scale=0.725,clip=]{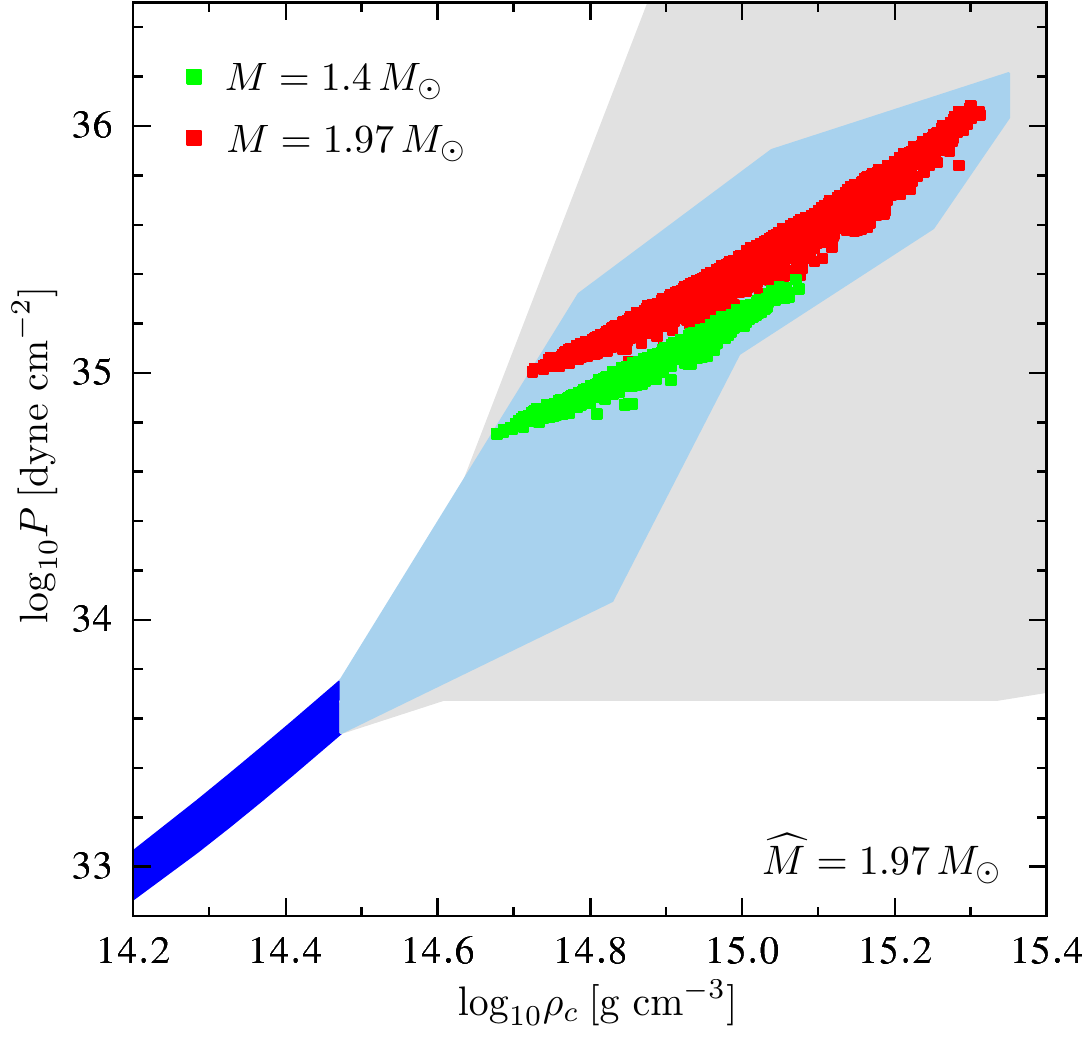}
\hspace{2mm}
\includegraphics[scale=0.725,clip=]{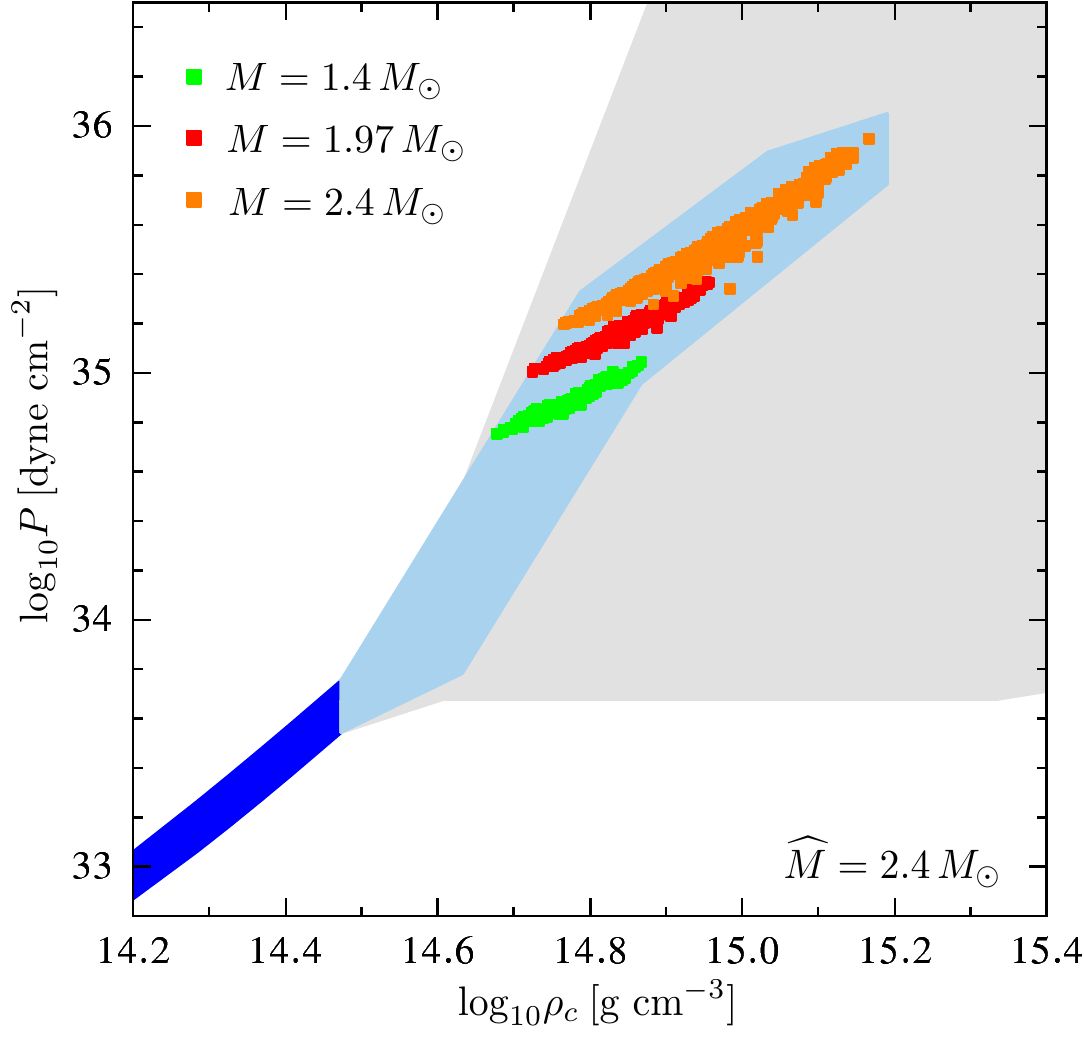}
\caption{(Color online) Central densities $\rho_c$ of neutron stars
with  masses $1.4 \, M_{\odot}$ (green~points), $1.97 \, M_{\odot}$
(red points), and $2.4 \, M_{\odot}$ (orange points) based on the 
uncertainty bands of Figure~\ref{prho_extension} for the two
cases $\widehat{M} = 1.97 \, M_\odot$ (left panel)
and $\widehat{M} = 2.4 \, M_\odot$ (right panel).\label{prho_core}}
\end{figure}

\begin{table}[t!]
\begin{center}
\begin{tabular}{l|cc|cc}
& \multicolumn{2}{c|}{$\widehat{M} = 1.97 \, M_{\odot}$} 
& \multicolumn{2}{c}{$\widehat{M} = 2.4 \, M_{\odot}$} \\
& min & max & min & max \\
\hline
$\rho_{c}/\rho_0$ ($1.4 \,M_{\odot}$)  & 1.8 & 4.4 & 1.8 & 2.7 \\
$\rho_{c}/\rho_0$ ($1.97 \,M_{\odot}$) & 2.0 & 7.6 & 2.0 & 3.4 \\
$\rho_{c}/\rho_0$ ($2.4 \,M_{\odot}$)  &     &     & 2.2 & 5.4 \\
\end{tabular}
\end{center}
\caption{Minimal and maximal central densities $\rho_c$ (in units of
the saturation density $\rho_0$) of the neutron stars shown by the
points in Figure~\ref{prho_core}.\label{tab:core_dens}}
\end{table}

Figure~\ref{prho_core} shows the central densities of neutron stars with
masses $1.4 \, M_{\odot}$, $1.97 \, M_{\odot}$, and $2.4 \, M_{\odot}$
based on the uncertainty bands of Figure~\ref{prho_extension}. The
results for the minimal and maximal central densities are given in
Table~\ref{tab:core_dens}. Since stiff EOSs along the upper limit of
the pressure uncertainty band are not sensitive to the mass
constraint, the minimal central densities are identical for the 
$\widehat{M}= 1.97 \, M_\odot$ and $\widehat{M} = 2.4 \, M_\odot$ cases.

\begin{figure}[t]
\begin{center}
\includegraphics[scale=0.84,clip=]{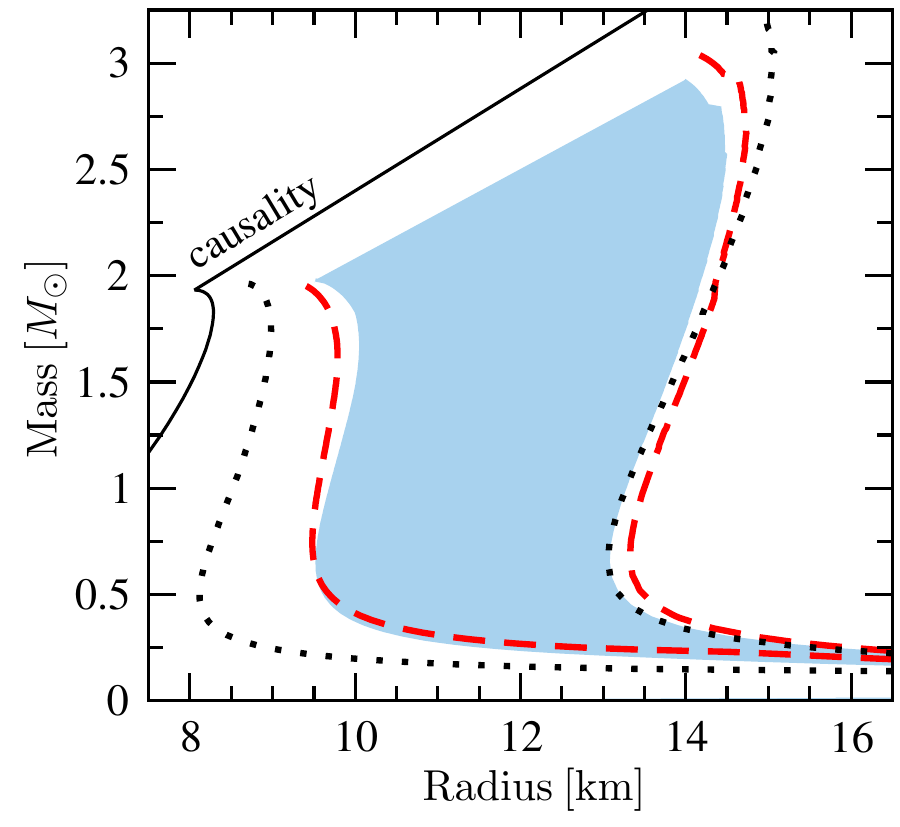}
\hspace{2mm}
\includegraphics[scale=0.84,clip=]{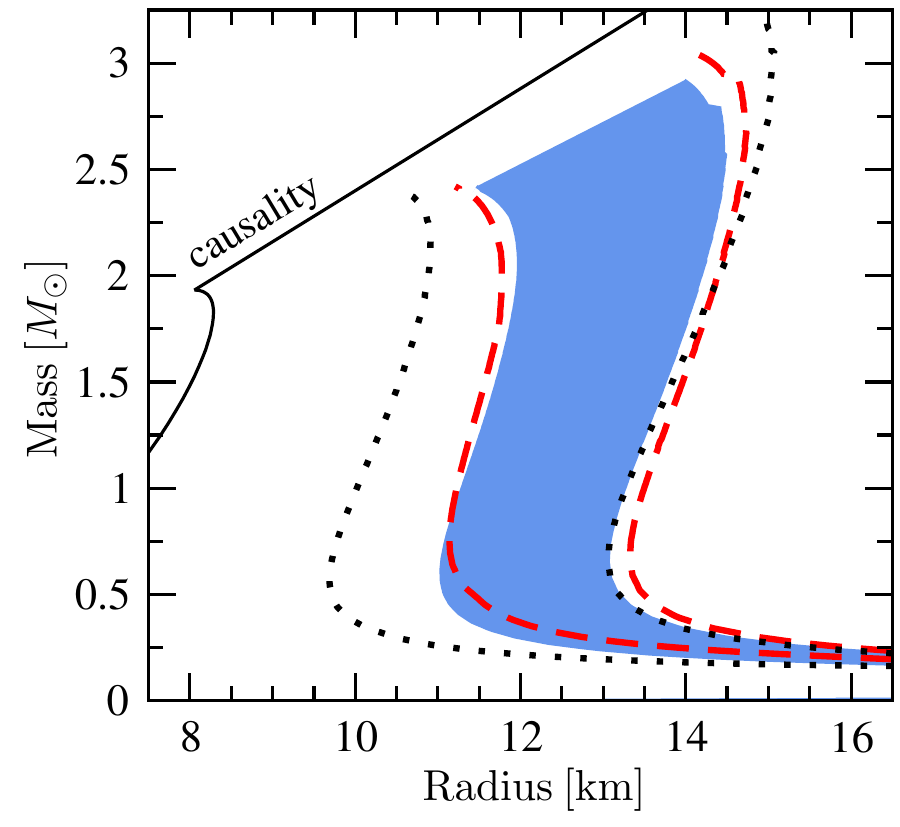}
\end{center}
\caption{(Color online) Neutron star mass-radius constraints based
on the uncertainty bands for the EOS of Figure~\ref{prho_extension}
for $\widehat{M} = 1.97 \, M_{\odot}$ (left panel) and $\widehat{M} = 2.4
\, M_{\odot}$ (right panel). The blue regions give the radius
constraints based on the neutron matter results with
renormalization-group-evolved interactions. The red dashed lines mark the band
without renormalization-group evolution (see
Figure~\ref{EN_nobeta}) and the thick dotted lines are for the allowed
EOS range with causal extension (see Figure~\ref{pepsilon}). The
solid line gives the limit~\citep{Lattimer}.\label{MvsR}}
\end{figure}

The uncertainty bands for the EOS directly translate into constraints
for the radii of neutron stars. In Figure~\ref{MvsR}, we present the
radius constraints obtained from the EOS bands of
Figure~\ref{prho_extension}, for the two cases $\widehat{M} = 1.97 \,
M_{\odot}$ (left panel) and $\widehat{M} = 2.4 \, M_{\odot}$ (right
panel). The mass-radius relationships for the individual EOSs were
obtained by solving the Tolman-Oppenheimer-Volkov equations and from
these an envelope was constructed. The blue regions in
Figure~\ref{MvsR} show the radius constraints based on the neutron
matter results with renormalization-group-evolved interactions. In addition, we
show results without renormalization-group evolution (see
Figure~\ref{EN_nobeta}) and for the allowed EOS range with causal
extension (see Figure~\ref{pepsilon}). The results with causal extension
are also based on renormalization-group-evolved interactions.

\begin{table}[t]
\begin{center}
\begin{tabular}{l|ccc|ccc}
& \multicolumn{3}{c|}{$\widehat{M} = 1.97 \, M_{\odot}$} 
& \multicolumn{3}{c}{$\widehat{M} = 2.4 \, M_{\odot}$} \\
& evolved & unevolved & causal & evolved & unevolved & causal \\
\hline
$R_{\rm min}$ ($1.4 \, M_{\odot}$) & 10.0 & 9.7  & 8.8  & 11.6 & 11.5 & 10.4 \\
$R_{\rm max}$ ($1.4 \, M_{\odot}$) & 13.7 & 13.9 & 13.7 & 13.7 & 13.9 & 13.7 \\
\hline
$R_{\rm min}$ ($1.97 \, M_{\odot}$) & 9.6  & 9.3  & 8.6  & 12.0 & 11.8 & 10.9 \\
$R_{\rm max}$ ($1.97 \, M_{\odot}$) & 14.2 & 14.4 & 14.4 & 14.2 & 14.4 & 14.4
\end{tabular}
\end{center}
\caption{Radius constraints for a $1.4 \, M_{\odot}$ neutron star
and for the heaviest known neutron star with $M = 1.97 \, M_{\odot}$ based
on the results of Figure~\ref{MvsR} (see text for details on the column
labels). The rows give the minimum and maximum radii.\label{tab:radii}}
\end{table}

The predicted radius ranges are given in Table~\ref{tab:radii} for
a $1.4 \, M_{\odot}$ neutron star and for the heaviest known
neutron star, with $M = 1.97 \, M_{\odot}$. The maximal radius is very
robust and essentially independent of the details of the neutron
matter calculation, the use of causal extensions, and the mass
constraints. This can be understood intuitively. Very stiff EOSs,
which determine the maximal radius constraint, lead also to large
neutron star masses. Hence, the constraints $\widehat{M} = 1.97 \,
M_{\odot}$ and $\widehat{M} = 2.4 \, M_{\odot}$ are always
fulfilled. Furthermore, the central densities of such neutron stars are
typically rather low (see Figure~\ref{prho_core}). For the stiffest
EOSs we find the central density to be $\rho_c \approx 1.8 \, \rho_0$
for an $M = 1.4 \, M_{\odot}$ neutron star, which is typical for many
observed neutron stars (see Table~\ref{tab:core_dens}), a density
region which is still rather well constrained by chiral EFT
interactions~\citep{nm,Weise}. A $1.4 \, M_{\odot}$ neutron star with
a radius significantly larger than $R = 13.9 \, {\rm km}$ would
therefore be incompatible with constraints from chiral EFT
interactions.

The minimal radius is more sensitive to details of the EOS at higher
densities and therefore less well constrained. The limits based on
evolved and unevolved nuclear interactions agree well, except for very
light neutron stars which are more sensitive to small differences in
the low-density part of the EOS. In addition, the minimal radius
strongly depends on the mass constraint. This implies that the lower
limit of theoretical neutron star radii can be systematically improved
by future observations of heavier neutron stars. Since the maximal
mass of a neutron star based on soft EOSs can be increased by causal
extensions (see the discussion of Figure~\ref{pepsilon}), we find a
reduction of the minimal radius for EOSs by about 1~km with causal
extensions, as shown by the difference between the blue bands and the
dotted lines. This is consistent with the lower radius limits
of~\cite{Lattimer}, giving $R_{\rm min}(1.4 \, M_{\odot}) = 8.1 \,
{\rm km}$ and $9.1 \, {\rm km}$ for $\widehat{M} = 1.97 \, M_{\odot}$
and $\widehat{M} = 2.4 \, M_{\odot}$, respectively.

\section{Representative equations of state}
\label{sec:repEOS}

\begin{figure}[t]
\begin{center}
\includegraphics[scale=0.875,clip=]{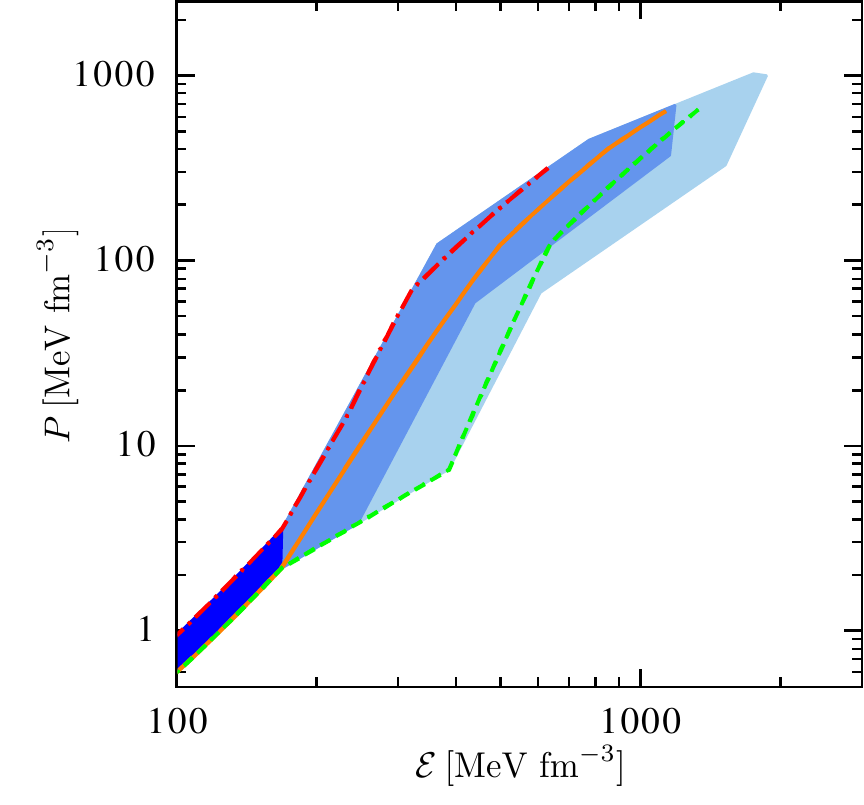}
\hspace{4mm}
\includegraphics[scale=0.875,clip=]{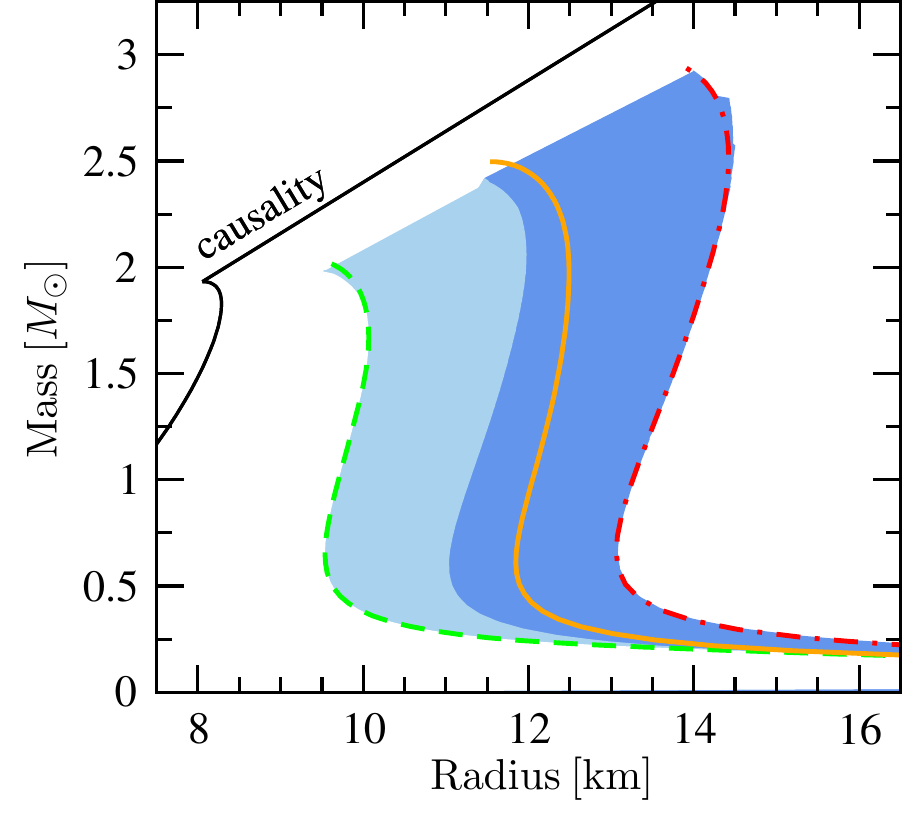}
\end{center}
\caption{(Color online) Left panel: Pressure $P$ of neutron star matter
as a function of energy density ${\mathcal E}$ for three representative
EOSs in comparison with the uncertainty bands of Figure~\ref{prho_extension}.
Right panel: The corresponding neutron star mass-radius results for
the three EOSs and the uncertainty bands of Figure~\ref{MvsR}.\label{repEOS}}
\end{figure}

The uncertainty bands for the EOS are an envelope of many individual
EOSs. Therefore, the limits of the uncertainty bands are not given by
one EOS in general~(see Figure~\ref{prho_extension}). In order to
explore the EOS bands in astrophysical applications, we present three
representative EOSs that probe the central region and the extremes of
the uncertainty band for $\widehat{M} = 1.97 \, M_{\odot}$:

\begin{itemize}
\item[1.)] a soft EOS (green dashed line in Figure~\ref{repEOS}), which agrees
well with the lower limit of the EOS band for low and medium
densities. This EOS describes excellently the minimal radius over the
entire mass range (see the right panel of Figure~\ref{repEOS}).  At
higher densities, the pressure then reaches values above the lower
limit of the band, as a consequence of the mass constraint. The
polytropic parameters of this EOS are $\Gamma_1 = 1.5$, $\rho_{12} =
2.5 \, \rho_0$, $\Gamma_2 = 6.0$, $\rho_{23} = 4.0 \, \rho_0$,
$\Gamma_3 = 3.0$, and $\rho_{\rm max} \approx 7.0 \, \rho_0$.

\item[2.)] an intermediate EOS (orange solid line in Figure~\ref{repEOS}), which 
is located in the central region of the EOS band and is also
consistent with the $\widehat{M} = 2.4 \, M_{\odot}$ EOS band over the
whole density range. The radius of neutron stars for this EOS is about
$12 \, {\rm km}$ for typical masses, and lies in the center of the
predicted radius range. The polytropic parameters of this EOS are
$\Gamma_1 = 4.0$, $\rho_{12} = 3.0 \, \rho_0$, $\Gamma_2 = 3.0$,
$\rho_{23} = 4.5 \, \rho_0$, $\Gamma_3 = 2.5$, and $\rho_{\rm max}
\approx 5.4 \, \rho_0$.

\item[3.)] a stiff EOS (red dash-dotted line in Figure~\ref{repEOS}), which 
follows closely the upper limit of the EOS band until the causal limit
is reached. This EOS gives the largest neutron star radii consistent
with the constraints. The polytropic parameters of this EOS are
$\Gamma_1 = 4.5$, $\rho_{12} = 1.5 \, \rho_0$, $\Gamma_2 = 5.5$,
$\rho_{23} = 2.0 \, \rho_0$, $\Gamma_3 = 3.0$, and $\rho_{\rm max} =
3.3 \, \rho_0$.
\end{itemize}

\noindent
Detailed values of the pressure, energy density, and energy per
particle for these three EOSs as a function of number
density and mass density are listed in Tables~\ref{tab:repEOS1}
and~\ref{tab:repEOS2}, with the BPS crust EOS used at low densities in
Table~\ref{tab:repEOS3}. In addition, Tables~\ref{tab:repEOS1}
and~\ref{tab:repEOS2} include values for the neutron star radius and
mass at that central density.

\section{Conclusions and outlook}

Recent advances in nuclear theory combined with new astrophysical
observations have systematically tightened the constraints on the EOS
of neutron-rich matter over a wide range of densities. In this paper,
we have presented constraints for the EOS and for the structure of
neutron stars based on microscopic neutron matter calculations with
chiral EFT interactions. At nuclear saturation density and below, the
uncertainties are dominated by uncertainties in 3N forces. In the
future, these uncertainties can be reduced by improved constraints on
the low-energy couplings entering 3N forces and by systematic
improvements in chiral EFT that incorporate higher-order many-body
forces~\citep{N3LO,N3LO_long} and $\Delta$ degrees of freedom
explicitly. In addition, reduced uncertainties are expected when
renormalization-group-evolved 3N
forces~\citep{3N_evolution_mom,nm_evolved} and nonperturbative
many-body calculations are employed.

The uncertainties of the neutron matter calculations directly
translate into constraints for the symmetry energy $S_v$ and its
density derivative $L$, which is related to the pressure of neutron
matter. On the assumptions that the interaction energy depends
quadratically on the neutron excess, this leads to $S_v = 29.7 - 33.5
\mev$ and $L = 32.4 - 57.0 \mev$. These ranges provides very tight
constraints and are consistent within uncertainties with different
empirical extractions of these parameters (see
Figure~\ref{symcor}). In addition, the neutron matter calculations
predict the crust-core transition density to be in the range $(0.475 -
0.55) \, n_0$ when Coulomb and density gradient corrections are included.

We extended the EOS to higher densities by employing a general
piecewise polytropic extrapolation, as illustrated in
Figure~\ref{fig:polytropic}.  This led to a very large number of EOSs,
which cover a wide pressure range, including the possibility of a soft
phase transition region. {}From the individual EOSs we selected those
that remain causal at all densities relevant in neutron stars and are
able to support a neutron star with mass $1.97 \, M_{\odot}$, the
heaviest known neutron star~\citep{Demorest}. We also considered a
second case where the EOS supports a neutron star with mass $2.4 \,
M_{\odot}$. Note that the constraints are in accord with our previous
work~\citep{Kai}, where we used only two polytropes.  Combined with
the microscopic neutron matter calculations, this provides tight
constraints on the nuclear EOS at sub- and supranuclear densities, and
rules out many model EOSs, as demonstrated by
Figure~\ref{prho_compare}.

Based on the allowed EOS band, we predicted ranges for the radii
and central densities of neutron stars (see Figures~\ref{prho_core}
and~\ref{MvsR} as well as Tables~\ref{tab:core_dens}
and~\ref{tab:radii}). For the constraint $\widehat{M} = 1.97 \,
M_{\odot}$, the radius of a $1.4 \, M_{\odot}$ neutron star is
predicted to be $9.7 - 13.9 \, {\rm km}$ (based on
renormalization-group-evolved or unevolved chiral EFT interactions)
with central densities up to $4.4 \, \rho_0$. Note that the maximum
radius is determined by the causality constraint, so that the
discovery of a heavier neutron star only affects the minimum radius.
If a $2.4 \, M_{\odot}$ neutron star were discovered, this would imply
a radius range $11.5 - 13.9 \, {\rm km}$ for a $1.4 \, M_{\odot}$
neutron star, with central densities up to only $2.7 \, \rho_0$.

For use in astrophysical simulations, we have constructed three
representative EOSs (soft, intermediate, and stiff), consistent with
the constraints from nuclear physics and observation. In addition to
the nuclear physics improvements mentioned above and the measurement
of heavier neutron stars, information on the radii of neutron stars
will significantly tighten the EOS band at high densities. Currently,
observational radius limits are often conflicting. Studies of
five photospheric radius expansion X-ray bursts by
\citet{Steiner1,Steiner2} are in agreement with the present analysis,
but other analyses \citep{Ozel} of two of them (EXO1745-248 and
4U1820-30) indicate radii that are too small to be compatible with our
results. On the other hand, a study of the burster 4U1724-307
\citep{Suleimanov} yields a radius too large to be compatible. Using
phase modeling of pulsar profiles, \citet{Bogdanov} found a compatible
$3\sigma$ lower limit to the radius of pulsar PSR J0437-4715 of
11.1~km, assuming the $1.56 M_\odot$ lower limit for its mass found by
\citet{Verbiest}. However, \citet{Hambaryan} found
$R=(10-12)(M/M_\odot) \, {\rm km}$ for the pulsar RBS 1223 using the
same technique, which is compatible only if $M<1.4 M_\odot$. 
Also, using pulse profile modeling, \citet{Leahy11} found radii
for the sources SAX J1808.4-3658, XTE J1807-294 and XTE J1814-338
which are compatible with our results. A recent analysis \citep{Guillot} of 5 quiescent
neutron stars in low-mass X-ray binaries in globular clusters, in
which it was assumed that the stars all had the same radius,
determined the radius to be $R = 9.1^{+1.3}_{-1.5} \, {\rm km}$ to
90\% confidence, which is marginally consistent with our
results. Additional radius measurements from the LOFT~\citep{LOFT} and
NICER~\citep{NICER} X-ray missions are therefore eagerly anticipated.

\section*{Acknowledgments}

We thank T.\ Kr\"uger and I.\ Tews for helpful discussions. 
This work was supported in part by the NSF under Grant No.~PHY--1002478,
the US Department of Energy under Grants DE-FG02-87ER40317 and 
DE-SC0008533 (SciDAC-3 NUCLEI project), the ESF 
AstroSim and CompStar networks, the Helmholtz Alliance Program of the 
Helmholtz Association, contract HA216/EMMI ``Extremes of Density and 
Temperature: Cosmic Matter in the Laboratory'', the DFG through Grant 
SFB 634, and by the ERC Grant No.~307986 STRONGINT.

\appendix

\section{Instability of the uniform phase}
\label{instability}

Here we discuss details of the density-gradient contributions to the
instability condition and also express the instability condition in
terms of the variables $n$ and $x$, as was done in \citet{LP2007},
rather than $n_n$ and $n_p$. The contribution of density gradients to
the energy density is expressed in the form~\citep{BBP}
\begin{equation}
{\mathcal E}_{\rm grad} = Q_{pp}(\nabla n_p)^2 + 2Q_{pn}(\nabla n_p)\cdot
(\nabla n_n) + Q_{nn}(\nabla n_n)^2 \,,
\end{equation}
where the coefficients $Q_{ij}$ are in general functions of the
neutron and proton densities, but will here be treated as constants.
We follow \citet{BBP} and take $Q_{np}=Q_{pn} = 2 Q_{nn} = 2 Q_{pp}$,
in which case the quantity $\beta$ of Section~\ref{sec:crust-core} is
given by
\begin{equation}
\beta = 2 Q_{nn} (1+4\zeta+\zeta^2) \,.
\end{equation}
Indeed, realistic Skyrme density functionals such as Ska and SkM$^*$
satisfy this approximate relationship among the $Q_{ij}$
values~\citep{LP2007}. It would be interesting to determine the
$Q_{ij}$ from modern energy density functionals~\citep{Erler1,Erler2}.

The quantity $Q$ may be determined either from the surface energy of
symmetric nuclear matter or from the surface thickness of symmetric
nuclei. For symmetric nuclear matter, the energy density is given by
\begin{equation}
{\mathcal E}_{\rm bulk} + {\mathcal E}_{\rm grad} = 
n (\epsilon_{\rm s}(n) + m c^2) + Q(\nabla n)^2 \,,
\end{equation}
where $n [\epsilon_{\rm s} (n) + m c^2] = {\mathcal E}(n,x=1/2)$,
$Q=3Q_{nn}/2=3Q_{np}/4$ and $m = (m_n + m_p)/2$. In the Thomas-Fermi
approximation and for symmetric nuclear matter the optimum density
profile for a plane surface is determined by minimizing the quantity
$\int dz \, [n (\epsilon_{\rm s}(n) - \mu_0) + Q (n')^2]$ for a surface
lying in the $xy$-plane~\citep{Bethe}. Here $\mu_0=-B$ is the chemical
potential of symmetric nuclear matter at the saturation density and
the prime denotes a derivative with respect to $z$. This leads to the
condition
\begin{equation}
n^\prime = -\sqrt{\frac{n[\epsilon_{\rm s}(n) - \mu_0]}{Q}} = 
-\sqrt{\frac{n_0T_0}{Q}} \, f(n/n_0) \,,
\end{equation}
when the nuclear matter is assumed to occupy the region with negative
$z$. This equation defines $f(n/n_0)$, and, on integration, determines
$n(z)$. Note that $f(n/n_0) \rightarrow 0$ for both $n \rightarrow 0$
and $n \rightarrow n_0$. The surface energy per unit area, the surface
tension $\sigma_0$, is then given by
\begin{equation}
\sigma_0 = 2 I_\sigma \, \sqrt{QT_0n_0^3} \,,
\end{equation}
where
\begin{equation}
I_\sigma = \int_0^1 f(u) \, du \approx 0.1696 \,,
\end{equation}
for the bulk energy per particle given by Equation~(\ref{Eskyrme})
with $\gamma=4/3$, $\alpha = 5.87$, and $\eta = 3.81$ [see details
after Equation~(\ref{Pskyrme})]. The $90\%-10\%$ surface thickness can
be expressed as
\begin{equation}
t_{90-10}=I_t \, \sqrt{\frac{Qn_0}{T_0}} \,,
\end{equation}
where
\begin{equation}
I_t= \int_{0.1}^{0.9}\frac{du}{f(u)} \approx 4.776 \,.
\end{equation}
For $Q=75 \, {\rm MeV} \, {\rm fm}^5$, one finds $\sigma_0 \approx
1.15 \, {\rm MeV} \, {\rm fm}^{-2}$ and $t_{90-10} \approx 2.7 \,
{\rm fm}$, both reasonable values although perhaps $5\%$ too large for
the surface thickness. Note that Hartree-Fock calculations of surface
energies can differ by about $10 \%$ from the Thomas-Fermi
approximation \citep{HFvsTF}.  However, we have no freedom to fit both
observables with the adopted functional for the bulk energy per
particle. With this value for $Q$, we find $Q_{np}=100 \, {\rm MeV} \,
{\rm fm}^5$ and $Q_{pp}=Q_{nn}=50 \, {\rm MeV} \, {\rm fm}^5$, which
are the values used in the calculations in
Section~\ref{sec:crust-core}.

We now comment briefly on the stability conditions when expressed in
terms of the variables $n$ and $x$~\citep{LP2007}, rather than $n_n$
and $n_p$. To take care of the constraints on the numbers of neutrons
and protons, it is convenient to work with the thermodynamic potential
\begin{equation}
\Xi = {\mathcal E} - \mu_n^0 n_n - \mu_p^0 n_p \,,
\label{Xi}
\end{equation}
where $\mu_n^0$ and $\mu_p^0$ are the chemical potentials in the
initial state. With this choice, the first order terms in an expansion
of $\Xi$ in powers of deviations of the densities from those in the
initial state vanish.  With $n_n=n(1-x)$ and $n_p=nx$, it follows from
Equation~(\ref{Xi}) that
\begin{equation}
\delta^2\Xi= \frac{1}{2} \frac{\partial^2\cal E}{\partial n^2} \,
(\delta n)^2\ + \left( \frac{\partial^2\cal E}{\partial n\partial x}
+\mu_n^0-\mu_p^0 \right) \, \delta n \delta x + \frac{1}{2} 
\frac{\partial^2\cal E}{\partial x^2} \, (\delta x)^2 \,,
\label{quadraticform}
\end{equation}
where the term with the chemical potentials comes from the nonlinear
dependence of the neutron and proton densities on $n$ and $x$. Since
the baryon pressure is given by $P = n^2 \partial ({\cal E}/n)/\partial
n$ and the difference of the neutron and proton chemical potentials
by $n \, (\mu_p-\mu_n)=\partial {\cal E}/\partial x$, it follows that
\begin{equation}
\frac{\partial^2{\cal E}}{\partial n^2} = \frac{1}{n} 
\frac{\partial P}{\partial n} \,, \quad
\frac{\partial^2{\cal E}}{\partial x\partial n} = \frac{1}{n} 
\frac{\partial P}{\partial x} - \mu_n+\mu_p \,, \quad {\rm and} \quad 
\frac{1}{2} \frac{\partial^2\cal E}{\partial x^2} = n 
\frac{\partial(\mu_p-\mu_n)}{\partial x} \,.
\end{equation}
Therefore, since in the coefficients we may put $\mu_n^0=\mu_n$ and
$\mu_p^0=\mu_p$, the quadratic form, Equation~(\ref{quadraticform}), may
be rewritten as
\begin{equation}
\delta^2\Xi = \frac1{2n} \frac{\partial P}{\partial n} \, (\delta n)^2
+ \frac{1}{n} \frac{\partial P}{\partial x} \, \delta n \delta x 
+ \frac{n}{2} \frac{\partial(\mu_p-\mu_n)}{\partial x} \, (\delta x)^2 \,.
\end{equation}
The conditions for the quadratic form to be positive definite are that
the diagonal terms be positive, 
\begin{equation}
\frac{\partial P}{\partial n}>0 \quad {\rm and} \quad 
\frac{\partial (\mu_p-\mu_n)}{\partial x}>0 \,,
\label{stab1}
\end{equation}  
and that the determinant of the quadratic form be positive~\citep{LP2007},
\begin{equation}
\frac{\partial P}{\partial n} \, \frac{\partial (\mu_p-\mu_n)}{\partial x}
- \left(\frac{\partial P}{ \partial x}\right)^2>0 \,.
\label{stab2}
\end{equation}  
If the condition~(\ref{stab2}) and one of the conditions~(\ref{stab1})
are satisfied, the other condition~(\ref{stab1}) is satisfied
automatically. If Equation~(\ref{stab1}) is satisfied, the
inequality~(\ref{stab2}) may be rewritten as
\begin{equation}
\frac{\partial^2 {\cal E}}{\partial n^2} - \left(
\frac{\partial P}{\partial x}\right)^2 \frac{\partial x}{\partial 
(\mu_p-\mu_n)}>0 \,.
\end{equation}
The first term may be regarded as a direct interaction between density
fluctuations, without changes in the composition, while the second
represents an induced interaction between density fluctuations due to
changes in the composition.  For the system to be stable, the total
interaction consisting of direct and induced contributions must be positive.

\section{Tables}
\label{tables}

\begin{deluxetable}{cc|ccccc|ccccc|ccccc}
\tabletypesize{\small}
\rotate
\tablewidth{21.1cm}
\linespread{0.9}
\setlength{\tabcolsep}{0.175cm}
\tablecaption{Numerical data for the three representative EOSs of
Section~\ref{sec:repEOS} as a function of number density $n/n_0$
or mass density $\rho$. The mass density $\rho$ times $c^2$, pressure $P$, and
energy density ${\mathcal E}$ are given in ${\rm MeV} \fmiq$. The
energy per nucleon $\epsilon$ is given in MeV. We also list the
neutron star radius $R$ in km and mass $M$ in units of $M_{\odot}$
at the central density $\rho$.\label{tab:repEOS1}}
%parameters:
%lower limit: $\Gamma_1 = 1.5$, $\rho_{12} = 2.5$, $\Gamma_2 = 6.0$, $\rho_{23} = 4.0$, $\Gamma_3 = 3.0$, $\rho_{max} = 7.05$,
%center: $\Gamma_1 = 4.0$, $\rho_{12} = 3.0$, $\Gamma_2 = 3.0$, $\rho_{23} = 4.5$, $\Gamma_3 = 2.5$, $\rho_{max} = 5.4$
%upper limit: $\Gamma_1 = 4.5$, $\rho_{12} = 1.5$, $\Gamma_2 = 5.5$, $\rho_{23} = 2.0$, $\Gamma_3 = 3.0$, $\rho_{max} = 3.3$
\startdata
& & \multicolumn{5}{c|}{soft} & \multicolumn{5}{c|}{intermediate} & \multicolumn{5}{c}{stiff} \\
\tableline
\tableline
$n/n_0$ & $\rho c^2$ & $P$ & ${\mathcal E}$ & $\epsilon$ & $R$ & $M$ & $P$ & ${\mathcal E}$ & $\epsilon$ & $R$ & $M$ & $P$ & ${\mathcal E}$ & $\epsilon$ & $R$ & $M$ \\
\tableline
0.5792 & 87.07 & 0.4470 & 87.90 & 8.920 & 3260 & 0.53 & 0.4470 & 87.90 & 8.920 & 3260 & 0.53 &   0.6960& 87.99 & 9.937 & 77.04 & 0.10 \\
0.7124 & 107.1 & 0.7162 & 108.2 & 10.06 & 248.8 & 0.10 & 0.7162 & 108.2 & 10.06 & 248.8 & 0.10 & 1.150 & 108.4 & 11.75 & 29.32 & 0.13 \\
0.7861 & 118.1 & 0.9094 & 119.5 & 10.73 & 85.17 & 0.10 & 0.9094 & 119.5 & 10.73 & 85.17 & 0.10 & 1.473 & 119.7 & 12.82 & 23.22 & 0.15 \\
0.8646 & 129.9 & 1.154 & 131.5 & 11.47 & 47.29 & 0.10 & 1.154 & 131.5 & 11.47 & 47.29 & 0.10 &   1.880 & 131.9 & 14.03 & 19.72 & 0.18 \\
0.9483 & 142.5 & 1.464 & 144.4 & 12.31 & 32.05 & 0.12 & 1.464 & 144.4 & 12.31 & 32.05 & 0.12 &   2.392 & 144.8 & 15.38 & 17.50 & 0.20 \\
1.0371 & 155.9 & 1.851 & 158.0 & 13.24 & 24.36 & 0.13 & 1.851 & 158.0 & 13.24 & 24.36 & 0.13 &   3.028 & 158.7 & 16.91 & 16.02 & 0.23 \\
1.1    & 165.3 & 2.163 & 167.8 & 13.94 & 21.15 & 0.14 & 2.163 & 167.8 & 13.94 & 21.15 & 0.14 &   3.542 & 168.5 & 18.04 & 15.29 & 0.26 \\ \hline
1.2    & 180.4 & 2.465 & 183.3 & 14.99 & 19.26 & 0.15 & 3.064 & 183.3 & 15.12 & 16.57 & 0.17 &   5.240 & 184.2 & 20.05 & 13.98 & 0.34 \\
1.3    & 195.4 & 2.780 & 198.8 & 16.04 & 18.00 & 0.16 & 4.220 & 198.9 & 16.57 & 14.23 & 0.22 &   7.512 & 200.1 & 22.57 & 13.33 & 0.45 \\
1.4    & 210.5 & 3.106 & 214.3 & 17.05 & 17.08 & 0.17 & 5.677 & 214.6 & 18.25 & 13.01 & 0.28 &   10.48 & 216.2 & 25.63 & 13.09 & 0.58 \\
1.5    & 225.5 & 3.445 & 229.8 & 18.03 & 16.36 & 0.17 & 7.481 & 230.3 & 20.19 & 12.37 & 0.34 &   14.30 & 232.5 & 29.28 & 13.07 & 0.73 \\
1.6    & 240.5 & 3.795 & 245.4 & 18.97 & 15.80 & 0.18 & 9.684 & 246.3 & 22.41 & 12.04 & 0.42 &   20.39 & 249.2 & 33.74 & 13.22 & 0.96 \\
1.7    & 255.6 & 4.157 & 261.0 & 19.88 & 15.32 & 0.18 & 12.34 & 262.3 & 24.93 & 11.89 & 0.51 &   28.47 & 266.3 & 39.30 & 13.47 & 1.22 \\
1.8    & 270.6 & 4.529 & 276.6 & 20.77 & 14.92 & 0.19 & 15.51 & 278.6 & 27.76 & 11.85 & 0.61 &   38.98 & 283.9 & 46.12 & 13.75 & 1.50 \\
1.9    & 285.6 & 4.911 & 292.2 & 21.63 & 14.57 & 0.19 & 19.25 & 295.0 & 30.92 & 11.87 & 0.71 &   52.49 & 302.2 & 54.41 & 14.01 & 1.79 \\
2.0    & 300.7 & 5.304 & 307.9 & 22.47 & 14.26 & 0.20 & 23.64 & 311.7 & 34.43 & 11.93 & 0.82 &   69.59 & 321.3 & 64.37 & 14.23 & 2.07 \\
2.1    & 315.7 & 5.707 & 323.5 & 23.28 & 13.99 & 0.20 & 28.73 & 328.6 & 38.31 & 12.00 & 0.94 &   80.56 & 341.1 & 75.52 & 14.32 & 2.21 \\
2.2    & 330.7 & 6.119 & 339.2 & 24.08 & 13.74 & 0.21 & 34.61 & 345.7 & 42.58 & 12.09 & 1.05 &   92.63 & 361.4 & 87.21 & 14.38 & 2.33 \\
2.3    & 345.8 & 6.541 & 354.9 & 24.87 & 13.52 & 0.21 & 41.35 & 363.2 & 47.26 & 12.17 & 1.17 &   105.8 & 382.4 & 99.44 & 14.41 & 2.44 \\
2.4    & 360.8 & 6.972 & 370.6 & 25.63 & 13.31 & 0.21 & 49.02 & 380.9 & 52.36 & 12.25 & 1.29 &   120.2 & 403.9 & 112.2 & 14.42 & 2.53 \\
2.5    & 375.8 & 7.413 & 386.4 & 26.38 & 13.12 & 0.22 & 57.72 & 399.0 & 57.91 & 12.32 & 1.41 &   135.9 & 426.0 & 125.5 & 14.41 & 2.61 \\
2.6    & 390.9 & 9.379 & 402.2 & 27.18 & 12.33 & 0.23 & 67.52 & 417.4 & 63.91 & 12.37 & 1.53 &   152.9 & 448.9 & 139.4 & 14.38 & 2.68 \\
2.7    & 405.9 & 11.76 & 418.0 & 28.12 & 11.52 & 0.26 & 78.53 & 436.3 & 70.40 & 12.42 & 1.64 &   171.2 & 472.3 & 153.8 & 14.33 & 2.74 \\
2.8    & 420.9 & 14.63 & 434.0 & 29.20 & 10.83 & 0.29 & 90.82 & 455.6 & 77.38 & 12.45 & 1.74 &   190.9 & 496.5 & 168.7 & 14.28 & 2.79 \\
2.9    & 436.0 & 18.06 & 450.1 & 30.46 & 10.31 & 0.33 & 104.5 & 475.3 & 84.89 & 12.48 & 1.84 &   212.1 & 521.5 & 184.2 & 14.22 & 2.83 \\
3.0    & 451.0 & 22.13 & 466.3 & 31.90 &  9.95 & 0.39 & 119.6 & 495.6 & 92.92 & 12.49 & 1.94 &   234.8 & 547.1 & 200.3 & 14.15 & 2.86 \\
3.1    & 466.0 & 26.94 & 482.7 & 33.54 &  9.73 & 0.45 & 132.0 & 516.3 & 101.3 & 12.49 & 2.00 &   259.1 & 573.6 & 216.8 & 14.07 & 2.89 \\
3.2    & 481.1 & 32.60 & 499.2 & 35.41 &  9.60 & 0.52 & 145.2 & 537.4 & 110.1 & 12.48 & 2.06 &   285.0 & 600.9 & 234.0 & 14.00 & 2.92 \\
3.3    & 496.1 & 39.21 & 515.9 & 37.53 &  9.54 & 0.61 & 159.3 & 559.0 & 119.1 & 12.47 & 2.12 &   312.6 & 629.0 & 251.6 & 13.91 & 2.94 \\
\enddata
\end{deluxetable}

\begin{deluxetable}{cc|ccccc|ccccc}
\tabletypesize{\small}
\tablewidth{14.25cm}
\linespread{0.9}
\setlength{\tabcolsep}{0.175cm}
\tablecaption{Continuation of Table~\ref{tab:repEOS1}. The maximal
density for the stiff EOS is reached already in Table~\ref{tab:repEOS1}.
\label{tab:repEOS2}}
%parameters:
%lower limit: $\Gamma_1 = 1.5$, $\rho_{12} = 2.5$, $\Gamma_2 = 6.0$, $\rho_{23} = 4.0$, $\Gamma_3 = 3.0$, $\rho_{max} = 7.05$,
%center: $\Gamma_1 = 4.0$, $\rho_{12} = 3.0$, $\Gamma_2 = 3.0$, $\rho_{23} = 4.5$, $\Gamma_3 = 2.5$, $\rho_{max} = 5.4$
%upper limit: $\Gamma_1 = 4.5$, $\rho_{12} = 1.5$, $\Gamma_2 = 5.5$, $\rho_{23} = 2.0$, $\Gamma_3 = 3.0$, $\rho_{max} = 3.3$
\startdata
& & \multicolumn{5}{c|}{soft} & \multicolumn{5}{c}{intermediate} \\
\tableline
\tableline
$n/n_0$ & $\rho c^2$ & $P$ & ${\mathcal E}$ & $\epsilon$ & $R$ & $M$ & $P$ & ${\mathcal E}$ & $\epsilon$ & $R$ & $M$ \\
\tableline
3.4 & 511.1 & 46.90 & 532.8 & 39.92 & 9.54  & 0.69 & 174.2 & 581.0 & 128.3 & 12.45 & 2.17 \\
3.5 & 526.2 & 55.81 & 550.0 & 42.61 & 9.57  & 0.79 & 190.0 & 603.4 & 137.9 & 12.42 & 2.21 \\
3.6 & 541.2 & 66.09 & 567.5 & 45.62 & 9.63  & 0.89 & 206.8 & 626.3 & 147.7 & 12.39 & 2.25 \\
3.7 & 556.2 & 77.90 & 585.2 & 48.99 & 9.70  & 1.00 & 224.5 & 649.7 & 157.9 & 12.35 & 2.28 \\
3.8 & 571.3 & 91.41 & 603.3 & 52.74 & 9.77  & 1.11 & 243.2 & 673.6 & 168.2 & 12.31 & 2.32 \\
3.9 & 586.3 & 106.8 & 621.8 & 56.92 & 9.85  & 1.21 & 262.9 & 698.0 & 178.9 & 12.26 & 2.35 \\
4.0 & 601.3 & 124.3 & 640.7 & 61.54 & 9.93  & 1.32 & 283.7 & 722.9 & 189.9 & 12.22 & 2.37 \\
4.1 & 616.4 & 133.9 & 659.9 & 66.46 & 9.96  & 1.38 & 305.5 & 748.3 & 201.1 & 12.17 & 2.39 \\
4.2 & 631.4 & 143.9 & 679.4 & 71.50 & 9.99  & 1.43 & 328.4 & 774.3 & 212.6 & 12.12 & 2.41 \\
4.3 & 646.4 & 154.5 & 699.2 & 76.66 & 10.01 & 1.47 & 352.4 & 800.8 & 224.3 & 12.06 & 2.43 \\
4.4 & 661.5 & 165.5 & 719.1 & 81.94 & 10.03 & 1.55 & 377.6 & 827.9 & 236.4 & 12.01 & 2.44 \\
4.5 & 676.5 & 177.0 & 739.4 & 87.33 & 10.05 & 1.56 & 403.9 & 855.6 & 248.7 & 11.96 & 2.46 \\
4.6 & 691.5 & 189.1 & 759.9 & 92.87 & 10.06 & 1.59 & 426.7 & 883.8 & 261.3 & 11.91 & 2.47 \\
4.7 & 706.6 & 201.7 & 780.6 & 98.52 & 10.06 & 1.63 & 450.3 & 912.6 & 273.9 & 11.86 & 2.47 \\
4.8 & 721.6 & 214.9 & 801.7 & 104.3 & 10.07 & 1.66 & 474.7 & 941.8 & 286.7 & 11.82 & 2.48 \\
4.9 & 736.6 & 228.6 & 823.0 & 110.2 & 10.06 & 1.69 & 499.8 & 971.6 & 299.7 & 11.77 & 2.49 \\
5.0 & 751.7 & 242.8 & 844.6 & 116.2 & 10.06 & 1.72 & 525.6 & 1002  & 312.8 & 11.72 & 2.49 \\
5.1 & 766.7 & 257.7 & 866.5 & 122.3 & 10.05 & 1.75 & 552.3 & 1033  & 326.0 & 11.67 & 2.49 \\
5.2 & 781.7 & 273.2 & 888.7 & 128.6 & 10.05 & 1.77 & 579.8 & 1064  & 339.3 & 11.63 & 2.49 \\
5.3 & 796.8 & 289.2 & 911.2 & 135.0 & 10.03 & 1.80 & 608.1 & 1096  & 352.8 & 11.58 & 2.50 \\
5.4 & 811.8 & 305.9 & 934.0 & 141.5 & 10.02 & 1.82 & 637.2 & 1128  & 366.4 & 11.53 & 2.50 \\
5.5 & 826.8 & 323.2 & 957.1 & 148.1 & 10.00 & 1.84 & & & & & \\
5.6 & 841.9 & 341.2 & 980.6 & 154.8 & 9.99  & 1.86 & & & & & \\
5.7 & 856.9 & 359.8 & 1004  & 161.7 & 9.97  & 1.88 & & & & & \\
5.8 & 871.9 & 379.1 & 1028  & 168.7 & 9.95  & 1.89 & & & & & \\
5.9 & 886.9 & 399.0 & 1052  & 175.8 & 9.93  & 1.91 & & & & & \\
6.0 & 902.0 & 419.7 & 1077  & 183.0 & 9.90  & 1.92 & & & & & \\
6.1 & 917.0 & 441.1 & 1102  & 190.3 & 9.88  & 1.94 & & & & & \\
6.2 & 932.1 & 463.1 & 1128  & 197.8 & 9.85  & 1.95 & & & & & \\
6.3 & 947.1 & 485.8 & 1154  & 205.4 & 9.83  & 1.96 & & & & & \\
6.4 & 962.1 & 509.3 & 1180  & 213.1 & 9.80  & 1.97 & & & & & \\
6.5 & 977.1 & 533.6 & 1206  & 220.9 & 9.77  & 1.98 & & & & & \\
6.6 & 992.2 & 558.6 & 1233  & 228.9 & 9.75  & 1.99 & & & & & \\
6.7 & 1007  & 584.4 & 1261  & 237.0 & 9.72  & 1.99 & & & & & \\
6.8 & 1022  & 611.0 & 1289  & 245.2 & 9.69  & 2.00 & & & & & \\
6.9 & 1037  & 638.4 & 1317  & 253.5 & 9.66  & 2.01 & & & & & \\
7.0 & 1052  & 666.5 & 1345  & 261.9 & 9.63  & 2.01 & & & & &
\enddata
\end{deluxetable}

\begin{deluxetable}{cc|cc|p{0.05cm}|cc|cc}
\tabletypesize{\small}
\tablewidth{17.65cm}
\linespread{0.9}
\setlength{\tabcolsep}{0.175cm}
\tablecaption{Numerical data of the BPS EOS for the outer 
crust~\citep{BPS,Vautherin}. The units are as in Table~\ref{tab:repEOS1}.
\label{tab:repEOS3}}
\startdata
$n/n_0$ & $\rho c^2$ & $P$ & ${\mathcal E}$ & & $n/n_0$ & $\rho c^2$ & $P$ & ${\mathcal E}$ \\
\cline{1-4}
\cline{6-9}
$2.956 \: 10^{-14}$ & $4.444 \: 10^{-12}$ & $6.303 \: 10^{-25}$ & $4.385 \: 10^{-12}$ & & $1.972 \: 10^{-4}$ & $2.964 \: 10^{-2}$ & $3.713 \: 10^{-5}$ & $2.931 \: 10^{-2}$ \\
$2.975 \: 10^{-14}$ & $4.472 \: 10^{-12}$ & $6.303 \: 10^{-24}$ & $4.407 \: 10^{-12}$ & & $2.482 \: 10^{-4}$ & $3.732 \: 10^{-2}$ & $5.048 \: 10^{-5}$ & $3.692 \: 10^{-2}$ \\
$3.069 \: 10^{-14}$ & $4.613 \: 10^{-12}$ & $6.303 \: 10^{-23}$ & $4.547 \: 10^{-12}$ & & $3.125 \: 10^{-4}$ & $4.698 \: 10^{-2}$ & $6.865 \: 10^{-5}$ & $4.649 \: 10^{-2}$ \\
$4.369 \: 10^{-14}$ & $6.568 \: 10^{-12}$ & $7.551 \: 10^{-22}$ & $6.472 \: 10^{-12}$ & & $3.934 \: 10^{-4}$ & $5.914 \: 10^{-2}$ & $9.330 \: 10^{-5}$ & $5.852 \: 10^{-2}$ \\
$6.187 \: 10^{-14}$ & $9.302 \: 10^{-12}$ & $8.737 \: 10^{-21}$ & $9.150 \: 10^{-12}$ & & $4.953 \: 10^{-4}$ & $7.445 \: 10^{-2}$ & $1.269 \: 10^{-4}$ & $7.376 \: 10^{-2}$ \\
$1.700 \: 10^{-13}$ & $2.556 \: 10^{-11}$ & $1.061 \: 10^{-19}$ & $2.516 \: 10^{-11}$ & & $6.235 \: 10^{-4}$ & $9.373 \: 10^{-2}$ & $1.621 \: 10^{-4}$ & $9.284 \: 10^{-2}$ \\
$7.937 \: 10^{-13}$ & $1.193 \: 10^{-10}$ & $3.632 \: 10^{-18}$ & $1.183 \: 10^{-10}$ & & $6.906 \: 10^{-4}$ & $1.038 \: 10^{-1}$ & $1.805 \: 10^{-4}$ & $1.029 \: 10^{-1}$ \\
$4.331 \: 10^{-12}$ & $6.511 \: 10^{-10}$ & $1.186 \: 10^{-16}$ & $6.416 \: 10^{-10}$ & & $7.850 \: 10^{-4}$ & $1.180 \: 10^{-1}$ & $2.053 \: 10^{-4}$ & $1.169 \: 10^{-1}$ \\
$3.934 \: 10^{-11}$ & $5.915 \: 10^{-9}$ & $6.081 \: 10^{-15}$ & $5.825 \: 10^{-9}$ & & $9.881 \: 10^{-4}$ & $1.485 \: 10^{-1}$ & $2.791 \: 10^{-4}$ & $1.473 \: 10^{-1}$ \\
$9.881 \: 10^{-11}$ & $1.485 \: 10^{-8}$ & $3.100 \: 10^{-14}$ & $1.463 \: 10^{-8}$ & & $1.244 \: 10^{-3}$ & $1.870 \: 10^{-1}$ & $3.630 \: 10^{-4}$ & $1.855 \: 10^{-1}$ \\
$2.483 \: 10^{-10}$ & $3.732 \: 10^{-8}$ & $1.517 \: 10^{-13}$ & $3.675 \: 10^{-8}$ & & $1.608 \: 10^{-3}$ & $2.417 \: 10^{-1}$ & $4.871 \: 10^{-4}$ & $2.398 \: 10^{-1}$ \\
$6.235 \: 10^{-10}$ & $9.373 \: 10^{-8}$ & $7.183 \: 10^{-13}$ & $9.228 \: 10^{-8}$ & & $1.669 \: 10^{-3}$ & $2.509 \: 10^{-1}$ & $4.924 \: 10^{-4}$ & $2.488 \: 10^{-1}$ \\
$1.566 \: 10^{-9}$ & $2.355 \: 10^{-7}$ & $3.286 \: 10^{-12}$ & $2.319 \: 10^{-7}$ & & $1.954 \: 10^{-3}$ & $2.937 \: 10^{-1}$ & $5.212 \: 10^{-4}$ & $2.917 \: 10^{-1}$ \\
$3.934 \: 10^{-9}$ & $5.914 \: 10^{-7}$ & $1.447 \: 10^{-11}$ & $5.825 \: 10^{-7}$ & & $2.469 \: 10^{-3}$ & $3.712 \: 10^{-1}$ & $5.678 \: 10^{-4}$ & $3.688 \: 10^{-1}$ \\
$9.881 \: 10^{-9}$ & $1.485 \: 10^{-6}$ & $6.088 \: 10^{-11}$ & $1.463 \: 10^{-6}$ & & $2.974 \: 10^{-3}$ & $4.471 \: 10^{-1}$ & $6.135 \: 10^{-4}$ & $4.443 \: 10^{-1}$ \\
$2.483 \: 10^{-8}$ & $3.732 \: 10^{-6}$ & $2.441 \: 10^{-10}$ & $3.676 \: 10^{-6}$ & & $3.633 \: 10^{-3}$ & $5.461 \: 10^{-1}$ & $6.759 \: 10^{-4}$ & $5.427 \: 10^{-1}$ \\
$3.125 \: 10^{-8}$ & $4.698 \: 10^{-6}$ & $3.282 \: 10^{-10}$ & $4.627 \: 10^{-6}$ & & $4.464 \: 10^{-3}$ & $6.711 \: 10^{-1}$ & $7.601 \: 10^{-4}$ & $6.673 \: 10^{-1}$ \\
$6.235 \: 10^{-8}$ & $9.373 \: 10^{-6}$ & $8.955 \: 10^{-10}$ & $9.233 \: 10^{-6}$ & & $5.491 \: 10^{-3}$ & $8.255 \: 10^{-1}$ & $8.731 \: 10^{-4}$ & $8.207 \: 10^{-1}$ \\
$1.244 \: 10^{-7}$ & $1.870 \: 10^{-5}$ & $2.392 \: 10^{-9}$ & $1.842 \: 10^{-5}$ & & $6.250 \: 10^{-3}$ & $9.396 \: 10^{-1}$ & $5.944 \: 10^{-4}$ & $9.390 \: 10^{-1}$ \\
$2.482 \: 10^{-7}$ & $3.732 \: 10^{-5}$ & $6.278 \: 10^{-9}$ & $3.676 \: 10^{-5}$ & & $2.500 \: 10^{-2}$ & $3.758$ & $5.799 \: 10^{-3}$ & $3.762$ \\
$4.953 \: 10^{-7}$ & $7.445 \: 10^{-5}$ & $1.625 \: 10^{-8}$ & $7.337 \: 10^{-5}$ & & $5.000 \: 10^{-2}$ & $7.517$ & $1.166 \: 10^{-2}$ & $7.530$ \\
$9.881 \: 10^{-7}$ & $1.485 \: 10^{-4}$ & $4.166 \: 10^{-8}$ & $1.464 \: 10^{-4}$ & & $7.500 \: 10^{-2}$ & $11.27$ & $2.085 \: 10^{-2}$ & $11.30$ \\
$1.244 \: 10^{-6}$ & $1.870 \: 10^{-4}$ & $5.453 \: 10^{-8}$ & $1.843 \: 10^{-4}$ & & $1.000 \: 10^{-1}$ & $15.03$ & $3.216 \: 10^{-2}$ & $15.08$ \\
$1.972 \: 10^{-6}$ & $2.964 \: 10^{-4}$ & $1.017 \: 10^{-7}$ & $2.922 \: 10^{-4}$ & & $1.250 \: 10^{-1}$ & $18.79$ & $4.515 \: 10^{-2}$ & $18.86$ \\
$3.125 \: 10^{-6}$ & $4.698 \: 10^{-4}$ & $1.890 \: 10^{-7}$ & $4.631 \: 10^{-4}$ & & $1.500 \: 10^{-1}$ & $22.55$ & $5.961 \: 10^{-2}$ & $22.64$ \\
$3.934 \: 10^{-6}$ & $5.914 \: 10^{-4}$ & $2.577 \: 10^{-7}$ & $5.830 \: 10^{-4}$ & & $1.750 \: 10^{-1}$ & $26.31$ & $7.544 \: 10^{-2}$ & $26.42$ \\
$4.952 \: 10^{-6}$ & $7.445 \: 10^{-4}$ & $3.143 \: 10^{-7}$ & $7.342 \: 10^{-4}$ & & $2.000 \: 10^{-1}$ & $30.07$ & $9.260 \: 10^{-2}$ & $30.21$ \\
$6.235 \: 10^{-6}$ & $9.373 \: 10^{-4}$ & $4.281 \: 10^{-7}$ & $9.245 \: 10^{-4}$ & & $2.250 \: 10^{-1}$ & $33.82$ & $1.110 \: 10^{-1}$ & $34.00$ \\
$9.881 \: 10^{-6}$ & $1.485 \: 10^{-3}$ & $7.938 \: 10^{-7}$ & $1.465 \: 10^{-3}$ & & $2.500 \: 10^{-1}$ & $37.58$ & $1.308 \: 10^{-1}$ & $37.79$ \\
$1.566 \: 10^{-5}$ & $2.355 \: 10^{-3}$ & $1.470 \: 10^{-6}$ & $2.323 \: 10^{-3}$ & & $2.750 \: 10^{-1}$ & $41.34$ & $1.518 \: 10^{-1}$ & $41.58$ \\
$2.482 \: 10^{-5}$ & $3.732 \: 10^{-3}$ & $2.722 \: 10^{-6}$ & $3.683 \: 10^{-3}$ & & $3.000 \: 10^{-1}$ & $45.10$ & $1.742 \: 10^{-1}$ & $45.38$ \\
$3.125 \: 10^{-5}$ & $4.698 \: 10^{-3}$ & $3.533 \: 10^{-6}$ & $4.637 \: 10^{-3}$ & & $3.250 \: 10^{-1}$ & $48.86$ & $1.980 \: 10^{-1}$ & $49.18$ \\
$3.934 \: 10^{-5}$ & $5.914 \: 10^{-3}$ & $4.807 \: 10^{-6}$ & $5.836 \: 10^{-3}$ & & $3.500 \: 10^{-1}$ & $52.62$ & $2.231 \: 10^{-1}$ & $52.98$ \\
$4.952 \: 10^{-5}$ & $7.445 \: 10^{-3}$ & $6.540 \: 10^{-6}$ & $7.353 \: 10^{-3}$ & & $3.750 \: 10^{-1}$ & $56.37$ & $2.497 \: 10^{-1}$ & $56.78$ \\
$6.235 \: 10^{-5}$ & $9.373 \: 10^{-3}$ & $8.893 \: 10^{-6}$ & $9.256 \: 10^{-3}$ & & $4.000 \: 10^{-1}$ & $60.13$ & $2.777 \: 10^{-1}$ & $60.58$ \\
$7.850 \: 10^{-5}$ & $1.180 \: 10^{-2}$ & $1.209 \: 10^{-5}$ & $1.166 \: 10^{-2}$ & & $4.250 \: 10^{-1}$ & $63.89$ & $3.073 \: 10^{-1}$ & $64.38$ \\
$9.881 \: 10^{-5}$ & $1.485 \: 10^{-2}$ & $1.562 \: 10^{-5}$ & $1.468 \: 10^{-2}$ & & $4.500 \: 10^{-1}$ & $67.65$ & $3.385 \: 10^{-1}$ & $68.19$ \\
$1.244 \: 10^{-4}$ & $1.870 \: 10^{-2}$ & $2.124 \: 10^{-5}$ & $1.848 \: 10^{-2}$ & & $4.750 \: 10^{-1}$ & $71.41$ & $3.713 \: 10^{-1}$ & $72.00$ \\
$1.566 \: 10^{-4}$ & $2.355 \: 10^{-2}$ & $2.888 \: 10^{-5}$ & $2.328 \: 10^{-2}$ & & $5.000 \: 10^{-1}$ & $75.17$ & $4.054 \: 10^{-1}$ & $75.81$
\enddata
\end{deluxetable}

\begin{thebibliography}{99}
%\bibitem[Akmal \& Pandharipande(1997)]{AP}
%Akmal, A.\ \& Pandharipande, V.\ R.\ 1997, Phys. Rev. C, 56, 2261

\bibitem[Akmal et al.(1998)]{APR}
Akmal, A., Pandharipande, V.\ R., \& Ravenhall, D.\ G. 1998, 
Phys.\ Rev.\ C, 58, 1804

\bibitem[Andersson et al.(2011)]{Anderson}
Andersson, N.\ et al.\ 2011, Gen.\ Rel.\ Grav., 43, 409

\bibitem[Bauswein \& Janka(2012)]{Bauswein}
Bauswein, A.\ \& Janka, H.-Th.\ 2012, Phys.\ Rev.\ Lett., 108, 011101 

\bibitem[Bauswein et al.(2012)]{Bauswein2}
Bauswein, A., Janka, H.-T., Hebeler, K., \& Schwenk A.\ 2012, 
Phys.\ Rev.\ D, 86, 063001  

\bibitem[Baym et al.(1971a)]{BBP}
Baym, G., Bethe, H. A., \& Pethick, C.\ J.\ 1971, Nucl. Phys. A, 175, 225

\bibitem[Baym et al.(1971b)]{BPS}
Baym, G., Pethick, C.\ J., \& Sutherland, P.\ 1971, Astrophys. J., 170, 299

\bibitem[Bethe(1968)]{Bethe}
Bethe, H.\ A.\ 1968, Phys.\ Rev.\ 167, 879

\bibitem[Bogdanov(2013)]{Bogdanov}
Bogdanov, S.\ 2013, Astrophys.\ J., 762, 96

\bibitem[Bogner et al.(2003)]{Vlowk}
Bogner, S.\ K., Kuo, T.\ T.\ S., \& Schwenk, A.\ 2003,
Phys.\ Rept., 386, 1

\bibitem[Bogner et al.(2007)]{smooth}
Bogner, S.\ K., Furnstahl, R.\ J., Ramanan, S., \& Schwenk, A.\ 2007, 
Nucl.\ Phys.\ A, 784, 79

\bibitem[Bogner et al.(2010)]{PPNP}
Bogner, S.\ K., Furnstahl, R.\ J., \& Schwenk, A.\ 2010, 
Prog.\ Part.\ Nucl.\ Phys., 65, 94

\bibitem[Chen et al.(2010)]{Sn}
Chen, L.-W.\ et al.\ 2010, Phys. Rev. C, 82, 024321

\bibitem[Demorest et al.(2010)]{Demorest}
Demorest, P.\ B., Pennucci, T., Ransom, S.\ M., Roberts, M.\ S.\ E., 
\& Hessels, J.\ W.\ T. 2010, Nature, 467, 1081

\bibitem[Entem \& Machleidt(2003)]{EM}
Entem, D.\ R.\ \& Machleidt, R.\ 2003, Phys.\ Rev.\ C, 68, 041001(R)

\bibitem[Epelbaum et al.(2005)]{EGM}
Epelbaum, E., Gl\"ockle, W., \& Mei{\ss}ner, U.-G.\ 2005, 
Nucl.\ Phys.\ A, 747, 362

\bibitem[Epelbaum et al.(2009)]{RMP}
Epelbaum, E., Hammer, H.-W., \& Mei{\ss}ner, U.-G.\ 2009, 
Rev. Mod. Phys., 81, 1773

\bibitem[Erler et al.(2012a)]{Erler1}
Erler, J., Birge, N., Kortelainen, M., Nazarewicz, N., Olsen, E., 
Perhac, A.\ M., \& Stoitsov, M.\ 2012, Nature, 486, 509

\bibitem[Erler et al.(2012b)]{Erler2}
Erler, J., Horowitz, C.\ J., Nazarewicz, N., Rafalski, M., \& Reinhard, 
P.-G.\ 2012, arXiv:1211.6292

\bibitem[Feroci et al.(2012)]{LOFT}
Feroci, M.\ et al.\ 2012, Experimental Astronomy, 34, 415

\bibitem[Gandolfi et al.(2012)]{Gandolfi}
Gandolfi, S., Carlson, J., \& Reddy, S.\ 2012, 
Phys.\ Rev.\ C, 85, 032801(R) 

\bibitem[Gendreau et al.(2012)]{NICER}
Gendreau, K.\ C., Arzoumanian, Z., \& Okajima, T.\ 2012,
Proc. SPIE, 8443, 844313

\bibitem[G\"uver et al.(2013)]{Guever13}
G\"uver, T., \& \"Ozel, F.\ 2013, Astrophys.\ J.\ Lett., 765, L1

\bibitem[Guillot et al.(2013)]{Guillot}
Guillot, S., Servillat, M., Webb, N.\ A., \& Rutledge, R.\ E.\ 2013,
arXiv:1302.0023

\bibitem[Hambaryan et al.(2011)]{Hambaryan}
Hambaryan, V., Suleimanov, V., Schwope, A.\ D., Neuhauser, R., Werner K., 
\& Potekhin, A.\ Y.\ 2011, Astronomy \& Astrophysics, 534, 74

\bibitem[Hammer et al.(2013)]{RMP3N}
Hammer, H.-W., Nogga, A., \& Schwenk, A.\ 2013,
Rev.\ Mod.\ Phys., 85, 197

\bibitem[Haensel et al.(2006)]{nstar_book}
Haensel, P., Potekhin, A.\ Y., \& Yakovlev, D.\ G.\ 2006,
Neutron Stars 1: Equation of State and Structure, Springer

\bibitem[Heiselberg \& Pandharipande(2000)]{HP}
Heiselberg, H.\ \& Pandharipande, V.\ 2000, 
Annu.\ Rev.\ Nucl.\ Part.\ Sci., 50, 481

\bibitem[Hebeler(2012)]{3N_evolution_mom}
Hebeler, K.\ 2012, Phys.\ Rev.\ C, 85, 021002(R)

\bibitem[Hebeler et al.(2011)]{nucmatt}
Hebeler, K., Bogner, S.\ K., Furnstahl, R.\ J., Nogga, A., \&
Schwenk, A. 2011, Phys.\ Rev.\ C, 83, 031301(R)

\bibitem[Hebeler \& Furnstahl(2013)]{nm_evolved}
Hebeler, K.\ \& Furnstahl R.\ J.\ 2013, Phys.\ Rev.\ C, 87, 031302(R)

\bibitem[Hebeler et al.(2010)]{Kai}
Hebeler, K., Lattimer, J.\ M., Pethick, C.\ J., \& Schwenk, A.\ 2010,
Phys.\ Rev.\ Lett., 105, 161102

\bibitem[Hebeler et al.(2007)]{NN_evolution}
Hebeler, K., Schwenk, A., \& Friman, B.\ 2007, Phys.\ Lett.\ B, 648, 176

\bibitem[Hebeler \& Schwenk(2010)]{nm}
Hebeler, K.\ \& Schwenk, A.\ 2010, Phys.\ Rev.\ C, 82, 014314

\bibitem[Kortelainen et al.(2010)]{masses}
Kortelainen, M.\ et al.\ 2010, Phys.\ Rev.\ C, 82, 024313 

\bibitem[Kr\"uger et al.(2013)]{N3LO_long}
Kr\"uger, T., Tews, I., Hebeler, K., \& Schwenk, A.\ 2013,
arXiv:1304.2212

\bibitem[Lackey et al.(2012)] {Lackey} 
Lackey, B.\ D., Kyutoku, K., Shibata, M., Brady, P.\ R., \& 
Friedman J.\ L.\ 2012, Phys.\ Rev.\ D 85, 044061

\bibitem[Lattimer(2012)]{Lattimer} 
Lattimer, J.\ M.\ 2012, Annu.\ Rev.\ Nucl.\ Part.\ Sci., 62, 485

\bibitem[Lattimer \& Lim(2012)]{LL}
Lattimer, J.\ M.\ \& Lim, Y. 2013, Astrophys.\ J., 771, 51

\bibitem[Lattimer \& Prakash(2001)]{LP} 
Lattimer, J.\ M.\ \& Prakash, M.\ 2001, Astrophys.\ J., 550, 426

\bibitem[Lattimer \& Prakash(2007)]{LP2007} 
Lattimer, J.\ M.\ \& Prakash, M.\ 2007, Phys.\ Rept., 442, 109

%\bibitem[Lattimer \& Prakash(2011)]{LP2011}
%Lattimer, J.\ M.\ \& Prakash, M. 2011, Festschrift in Honor of Gerald E. Brown, Editor: Sabine Lee, World Scientific, 275; arXiv1012.3208

\bibitem[Leahy et al.(2011)]{Leahy11}
Leahy, D.\ A., Morsink, S.\ M., \& Chou, Y.\ 2011, Astrophy.\ J., 742, 17

%\bibitem[Lorenz et al.(1993)]{NS_crust}
%Lorenz, C.\ P., Ravenhall, D.\ G., \& Pethick, C.\ J.\ 1993,
%Phys.\ Rev.\ Lett., 70, 379

\bibitem[Mei{\ss}ner(2006)]{Meissner_private}
Mei{\ss}ner, U.-G.\ 2006, private communication

\bibitem[Negele \& Vautherin(1973)]{Vautherin}
Negele, J.\ W.\ \& Vautherin, D.\ 1973, Nucl.\ Phys.\ A, 207, 298

\bibitem[\"Ozel et al.(2010)]{Ozel}
\"Ozel, F., Baym G., \& G\"uver, T.\ 2010, Phys.\ Rev.\ D, 82, 101301

\bibitem[\"Ozel et al.(2012)]{Ozel12}
\"Ozel, F., Gould, A., \& G\"uver, T.\ 2012, Astrophys.\ J., 748, 5

%\bibitem[\"Ozel(2013)]{Ozel2}
%\"Ozel, F.\ 2013, Rep.\ Prog.\ Phys., 76, 01690

\bibitem[Pethick et al.(1995)]{Chris_instability}
Pethick, C.\ J., Ravenhall D.\ G., \& Lorenz C.\ P.\ 1995,
Nucl.\ Phys.\ A, 584, 675

\bibitem[Ravenhall et al.(1972)]{HFvsTF}
Ravenhall, D.\ G., Bennett, C.\ D., \& Pethick, C.\ J.\ 1972,
Phys.\ Rev.\ Lett., 28, 978

\bibitem[Read et al.(2009)]{poly}
Read, J.\ S., Lackey, B.\ D., Owen, B.\ J., \& Friedman, J.\ L.\ 2009,
Phys.\ Rev.\ D, 79, 124032

\bibitem[Rentmeester et al.(2003)]{Rentmeester}
Rentmeester, M.\ C.\ M., Timmermans, R.\ G.\ E., \& de Swart, J.\ J.\ 2003,
Phys.\ Rev.\ C, 67, 044001

\bibitem[Steiner et al.(2010)]{Steiner1}
Steiner, A.\ W., Lattimer, J.\ M., \& Brown, E.\ F.\ 2010,
Astrophys.\ J., 722, 33

\bibitem[Steiner et al.(2013)]{Steiner2}
Steiner, A.\ W., Lattimer, J.\ M., \& Brown, E.\ F.\ 2013,
Astrophys.\ J.\ Lett., 765, L5

\bibitem[Suleimanov et al.(2011)]{Suleimanov}
Suleimanov, V.\, Poutanen, J., Revnivtsev M., \& Werner K.\ 2011,
Astrophys.\ J., 742, 122

\bibitem[Tamii et al.(2011)]{Tamii}
Tamii, A.\ et al.\ 2011, Phys.\ Rev.\ Lett., 107, 062502

\bibitem[Tews et al.(2013)]{N3LO}
Tews, I., Kr\"uger, T., Hebeler, K., \& Schwenk, A.\ 2013, 
Phys.\ Rev.\ Lett., 110, 032504

\bibitem[Trippa et al.(2008)]{GDR}
Trippa, L., Col{\`o}, G., \& Vigezzi, E.\ 2008,
Phys.\ Rev.\ C, 77, 061304

\bibitem[Tsang et al.(2009)]{HIC}
Tsang, M.\ B.\ et al.\ 2009, Phys.\ Rev.\ Lett., 102, 122701

\bibitem[van Kerkwijk et al.(2011)]{Kerkwijk}
van Kerkwijk, M.\ H., Breton, R.\ P., \& Kulkarni, S.\ R.\ 2011, 
Astrophys.\ J., 728, 95

\bibitem[Verbiest et al.(2007)]{Verbiest}
Verbiest, J.P.W., Bailes, M., van Straten, W., et al. 2008, Astrophys.\ J. 679, 675

\bibitem[Weinberg(1990)]{Weinberg1}
Weinberg, S.\ 1990, Phys.\ Lett.\ B, 251, 288

\bibitem[Weinberg(1991)]{Weinberg2}
Weinberg, S.\ 1991, Nucl.\ Phys.\ B, 363, 3

\bibitem[Weise(2012)]{Weise}
Weise, W.\ 2012, Prog.\ Part.\ Nucl.\ Phys., 67, 299
\end{thebibliography}
\end{document}